\documentclass[12pt,preprint]{aastex}
\usepackage{graphicx,color}
\begin{document}

\newcommand{\red}{\textcolor{red}}
\newcommand{\blue}{\textcolor{blue}}

\title{X-ray Diagnostics of Thermal Conditions of
 the Hot Plasmas\\ in the Centaurus Cluster}
 
\author{I. Takahashi\altaffilmark{1}, M. Kawaharada\altaffilmark{2},
K. Makishima\altaffilmark{1,2}, 
K. Matsushita\altaffilmark{3}, 
Y. Fukazawa\altaffilmark{4}, 
Y. Ikebe\altaffilmark{5}, 
T. Kitaguchi\altaffilmark{1}, M. Kokubun\altaffilmark{6},
K. Nakazawa\altaffilmark{1}, 
S. Okuyama\altaffilmark{1}, N. Ota\altaffilmark{3},
and T. Tamura\altaffilmark{6}
}
\affil{1: Department of Physics, University of Tokyo, 
7-3-1 Hongo, Bunkyo-ku, \\Tokyo 113-0011, Japan}
\affil{2: Institute of Physical and Chemical Research (RIKEN), 
2-1 Hirosawa, Wako, \\Saitama 351-0198, Japan}
\affil{3: Department of Physics, Tokyo University of Science,
    Kagurazaka, Shinjuku-ku,\\ Tokyo 162-8601, Japan}
\affil{4: Department of Physical Science, School of Science, Hiroshima University,\\
1-3-1 Kagamiyama, Higashi-Hiroshima, Hiroshima 739-8526}
\affil{5: National Museum of Emerging Science and Innovation,
 2-41 Aomi, Koto-ku, \\Tokyo 135-0064, Japan}
 \affil{6: Institute of Space and Astronautical Science, Japan
Aerospace Exploration Agency,\\ 2-1-1 Yoshinodai, Sagamihara, Kanagawa
229-8510, Japan}

\begin{abstract}

X-ray data of the Centaurus cluster,
obtained with {\it XMM-Newton} for 45 ksec, were analyzed.
Deprojected EPIC spectra from concentric thin shell regions
were reproduced  equally well
by a single-phase plasma emission model,
or by a two-phase model developed by {\it ASCA},
both incorporating cool (1.7--2.0 keV) and  hot ($\sim 4$ keV) 
plasma temperatures.
However,  EPIC spectra with higher statistics,
accumulated over 3-dimentional thick shell regions,
were reproduced better
by the two-phase model than by the singe-phase one.
Therefore,  hot and cool plasma phases are inferred to 
co-exist in the cluster core  region within $\sim 70$ kpc.
The iron and silicon abundances of the plasma 
were reconfirmed to increase significantly towards the center,
while that of oxygen was consistent with being radially constant.
The implied non-solar abundance ratios explains away 
the previously reported  excess X-ray absorption from the central region.
Although an  additional cool ($\sim 0.7$ keV) emission 
was detected within $\sim 20$ kpc of the center,
the RGS data gave tight upper limits on  any emission 
with a tempeartures below $\sim  0.5$ keV.
These results are compiled into a magnetosphere model,
which interprets the cool phase as confined 
within closed magnetic loops anchored to the cD galaxy.
When combined with so-called Rosner-Tucker-Vaiana mechanism
which applies to solar coronae,
this model can potentially explain basic properties of the cool phase,
including its temperature and thermal stability.
\end{abstract}


\section{INTRODUCTION}

Intra-cluster medium (ICM),
i.e., the hot plasma confined within the 
gravitational potential of clusters of galaxies,
constitutes the most dominant {\em known} form of baryons.
The ICM was thought to cool over the Hubble time
by emitting optically-thin thermal X-rays,
because its radiative cooling time is estimated to be
$\sim 10^8$ yr at the center of many ``cD clusters''
(those hosting cD galaxies at their centers).
This inspired so-called cooling flow (hereafter CF) 
hypothesis~\citep[e.g.][]{fabian94}.
The idea was apparently supported by several soft X-ray features
observed around cD galaxies, 
including general decreases of the ICM temperature,
strong excess X-ray surface brightness,
and  excess photoelectric absorption up to a few times $10^{21}$ cm$^{-2}$.

With the first imaging capability over a broad X-ray energy band up to 10~keV
and a much improved energy resolution than was available before,
{\it ASCA} \citep{asca} has provided a number of new results
that  altogether cast serious doubt upon the reality of CFs.
These include;
a shortage of the predicted cool gas in the Hydra-A cluster~\citep{ikebe97};
the presence of uncooled ICM in 3-dimensional core regions
of the Centaurus cluster~\citep{fuka94,ikebe99};
a central excess brightness of Abell~1795
that does not depend strongly on the X-ray energy~\citep{xu98};
hierarchical potential structures
in several galaxy groups~\citep{ikebe96,matsushita98};
and systematic differences in the ICM chemical composition
between the central and outer regions of cD clusters~\citep{fuka00}.
We compiled all these results in \citet[hereafter Paper~I]{max01},
and argued that the classical CF hypothesis needs a significant revision.
The {\it ASCA} suggestions have been reconfirmed and reinforced by 
{\it Chandra} and {\it XMM-Newton} \citep[e.g.][]{tamura01a,kaastra01,peterson01}.

Although the CF scenario in its original form is no longer considered valid, 
the ICM temperature of a cD cluster does decrease toward the center.
Furthermore, the deprojected radial ICM temperature profiles
of a fair number of clusters all reduce to 
a single ``universal'' profile~\citep{allen01,kaastra04},
which reach,  at the cluster center, a minimum value of
\begin{equation}
T_{\rm c} = (1/2 \sim 1/3) T_{\rm h}
\label{eq:TcTh}
\end{equation}
\citep{allen01,kaastra04}.
Here, $T_{\rm h}$ is the ICM temperature observed outside $\sim 100$ kpc.
It is yet to be explained how the intriguing universal temperature profile
and the scaling of eq.(\ref{eq:TcTh}) are produced,
and how they are related to mechanisms which suppress CFs.

The concept of the universal temperature profile is 
based on a ``single-phase'' (hereafter 1P) picture,
that the ICM at a given three-dimensional radius 
is represented by a single dominant temperature
which monotonically decreases toward the center.
In contrast, the original {\it ASCA} results on cD clusters~(Paper~I)
are based primarily on a ``two-phase'' (hereafter 2P) view;
the ICM consists of a ``hot phase'' and a ``cool phase'',
intermixed together,
with the volume filling factor of the latter increasing toward the center.
This view was at the beginning motivated by
the complex angular response of the {\it ASCA} telescopes,
but was reinforced {\it a posteriori} by the fact
that 2P fits to the {\it ASCA} spectra of a cluster 
generally give a pair of well-defined temperatures 
regardless of the two-dimensional radii used to extract
the signal photons~(e.g. Fukazawa et al.\ 1994; Paper~I).
In other words, the 2P modeling is likely 
to be more than a mere convention.

Generally, the hot-phase and cool-phase temperatures,
derived under the 2P assumption,
are respectively close to $T_{\rm c}$ and $T_{\rm h}$ of
eq.(\ref{eq:TcTh}) obtained through the 1P analysis.
Furthermore, eq.(\ref{eq:TcTh}) holds
for the 2P results as well~\citep{ikebe02}.
The 1P and 2P approaches are thus  consistent in the sense
that both yield essentially the same two characteristic temperatures,
$T_{\rm c}$ and $T_{\rm h}$.
Nevertheless, their physical implications are much different.
The 1P situation would require a fine tuning 
between the radiative cooling and the putative heating mechanism,
or  ``quasi-hydrostatic" gas cooling \citep{masai04}.
To realize the 2P condition, in contrast,
the cool phase must be thermally insulated from the hot phase,
and heated in a stable manner,
so that it should neither evaporate ~\citep{takahara},
nor collapse due to cooling.
Yet another possibility is
that the ICM temperature takes a range of values
even at a single radius~\citep{kaastra04};
this is to be called ``multi-phase'' picture.

In order to identify the CF-suppression mechanism,
it is of basic importance to clarify
which of the 1P and 2P pictures (or the multiphase view)
is closer to the reality.
In spite of its importance, 
there have been few attempts to address this issue.
We hence  analyze in the present paper
the {\it XMM-Newton} data of the Centaurus cluster.
This object is best suited to our  purpose,
because it harbors the very prominent cool component
and has extensively been studied with {\it ASCA}.
 Our strategy is to examine whether or not deprojected 
spectra of this cluster require the presence of multiple 
temperatures at each 3-dimensional radius.
The data will also allow us to reconfirm
the lack of emission with temperature below the value of eq.(\ref{eq:TcTh}).

In \S2, we briefly summarize previous X-ray results on the Centaurus cluster, 
and describe the {\it XMM-Newton} observations in \S3.
Section 4, which forms the core of the present paper,
is devoted to the description of data analysis and results.
The results are discussed in \S5,
followed by a summary in \S6.
Throughout the paper, the Hubble constant is expressed
as $H_0 = 72$~km~$^{-1}$~Mpc$^{-1}$.
The three-dimensional radius 
and the projected radius are denoted as $R$ and $r$, respectively.
Errors represent 90~\% confidence limits unless stated otherwise.

\section{PREVIOUS X-RAY RESULTS ON THE CENTAURUS CLUSTER}
Located at a redshift of $z$ = 0.0104~\citep{cenopt86,galz}
where $1'$ corresponds to 12.5 kpc,
the Centaurus cluster (Abell~3526) is one of 
the most well studied objects 
among nearby clusters.
It exhibits a roughly round shape,
and its cD galaxy, NGC~4696,
does not show significant nuclear activity.
In X-rays, the Centaurus cluster has been observed extensively,
with {\it Uhuru}~\citep{2ucatalog,4ucatalog},
{\it OSO8}~\citep{closo8}, 
{\it HEAO1}~\citep{cenheao1},
{\it Einstein}~\citep{ceneinstein}, 
{\it EXOSAT}~\citep{edge91a,edge91b},
and {\it Ginga}~\citep{cenginga}.
These observations  yielded an ICM temperature of $\sim 4$~keV,
except in the central region~($r \lesssim 6'$)
where a very prominent cool component 
and an excess brightness are observed.

\citet{cenrosat} studied the Centaurus cluster using the data from 
the {\it ROSAT} PSPC.
They found some evidence of metallicity increase 
and excess absorption at the center.
Through a deprojection analysis,
they also derived 3-dimensional properties of the ICM under the 1P modeling,
reporting that the ICM is cooled down to $\sim 1$~keV toward the center.

Using {\it ASCA} with the much improved energy resolution
and the wider energy band,  
\citet{fuka94} successfully resolved
both the Fe-K line and low-energy emission lines,
and detected hot and cool components
with a  temperature of $\sim 4$  and $\sim 1$ keV, respectively.
The strong metallicity increase toward the center was reconfirmed.
They also discovered that the 3-dimensional core region ($R< 5'$) of the cluster
is filled mostly with the hot  component,
whereas the cool component occupies a minor fraction of the volume.
Through a more detailed analysis of the {\it ASCA} data, 
\citet{ikebephd} showed that
the 1P density and temperature profiles derived with {\it ROSAT}~\citep{cenrosat}
underpredict the hard X-ray flux from the central region,
and hence fail to explain the 0.5--10~keV {\it ASCA} data.
\citet{ikebephd}  instead found 
that the {\it ASCA} spectra from various regions near the center
can be described adequately by the 2P formalism 
employing the two components identified by \citet{fuka94}.
This reinforced the presence of the hot phase in the three-dimensional core region,
and justified the 2P approach.

The 2P formalism on the Centaurus cluster
was further refined and reinforced by \citet{ikebe99}.
Incorporating two temperatures of 
$T_{\rm h} \sim 4$~keV and $T_{\rm c} \sim 1.4$~keV,
together with a central metallicity increase,
they successfully constructed a 3-dimensional ICM model
that can simultaneously explain the  {\it ROSAT} surface brightness
and the  {\it ASCA} annular spectra,
namely the best-quality data sets available at that time.
The excess absorption at the center,
clamed by \citet{fabian94asca} based on the {\it ASCA} data,
was not reconfirmed by \citet{ikebe99}.

Utilizing the unprecedented angular resolution of {\it Chandra},
\citet{cencxo} detected a ``plume'' like X-ray feature
near the center of the Centaurus cluster,
reconfirming a previous observation
with the {\it ROSAT} High Resolution Imager~\citep{plume}.
Within $10''$ of the center including  the ``plume'' region,
a cool plasma component with a temperature of $\sim 0.7$ keV 
was clearly detected  \citep{cencxo}.
Employing the 1P formalism,
\citet{cencxo} also derived the ICM temperature 
distribution near the center ($\sim 3'$),
but the field of view of {\it Chandra} was not wide enough
to determine the global temperature structure
over the entire ``cool'' region ($r \lesssim 6'$).

There are some indications 
that a subcluster centered on the elliptical galaxy NGC~4709
is merging with the main body.
This region exhibits
an excess X-ray surface brightness
and a temperature increase to $\sim 5$ keV~
\citep{churazov99,furusho01,molendi02}.
However, these effects, seen $\sim 15'$ off  the center,
are considered to be small
in regions which we analyze ($r<12'$).
In fact, using  {\it Suzaku}, \cite{ota07} showed 
that  the ICM within the central $\sim 12'$  
can be regarded as hydrostatic:
specifically, they set an upper limit of $\sim 1,400$ km s$^{-1}$
on any large-scale ($\sim 100$ kpc) bulk motion in the ICM
over this region,
and ruled out  a previously claimed detection of 
significant bulk motions \citep{dupke06}.

\section{OBSERVATIONS AND DATA REDUCTION}

\subsection{Observation}
The Centaurus cluster was observed with {\it XMM-Newton}
on 2002 January~3 for a gross exposure of 45 ksec.
The boresight was set to ($\alpha^{\rm 2000}$, $\delta^{\rm 2000}$) = 
$(12^{\rm h} 48^{\rm m} 49^{\rm s}.3, -41^\circ 18' 40''.0)$.
The EPIC PN and MOS were operated both in the full frame mode
with the thin filters~\citep{pn,mos},
while the RGS in spectroscopy mode~\citep{rgs}.

The EPIC dataset obtained in the same observation 
were already analyzed by \cite{matsushita07aap}
mainly for metallicity distributions.
The present paper emphasizes the temperature structure.
Similarly, \cite{cenrgs} utilized the RGS data from the same observation,
as well as those from an additional 110 ks observation.

\subsection{EPIC Data Reduction} \label{sub:EPIC}
\subsubsection{Data processing} \label{subsub:EPIC_process}
The EPIC data were extracted using the standard SAS software of version~5.4.1.
We selected events with 
{\it pattern} $\leq 4$ for PN and {\it pattern} $\leq 12$ for MOS,
with which most of the valid X-ray events are accepted.
According to instructions by the detector team,
we further discarded events with bad flags,
e.g., those out of the field of view and those detected at bad CCD pixels.

\begin{figure}\begin{center}
\epsscale{0.6}
\plotone{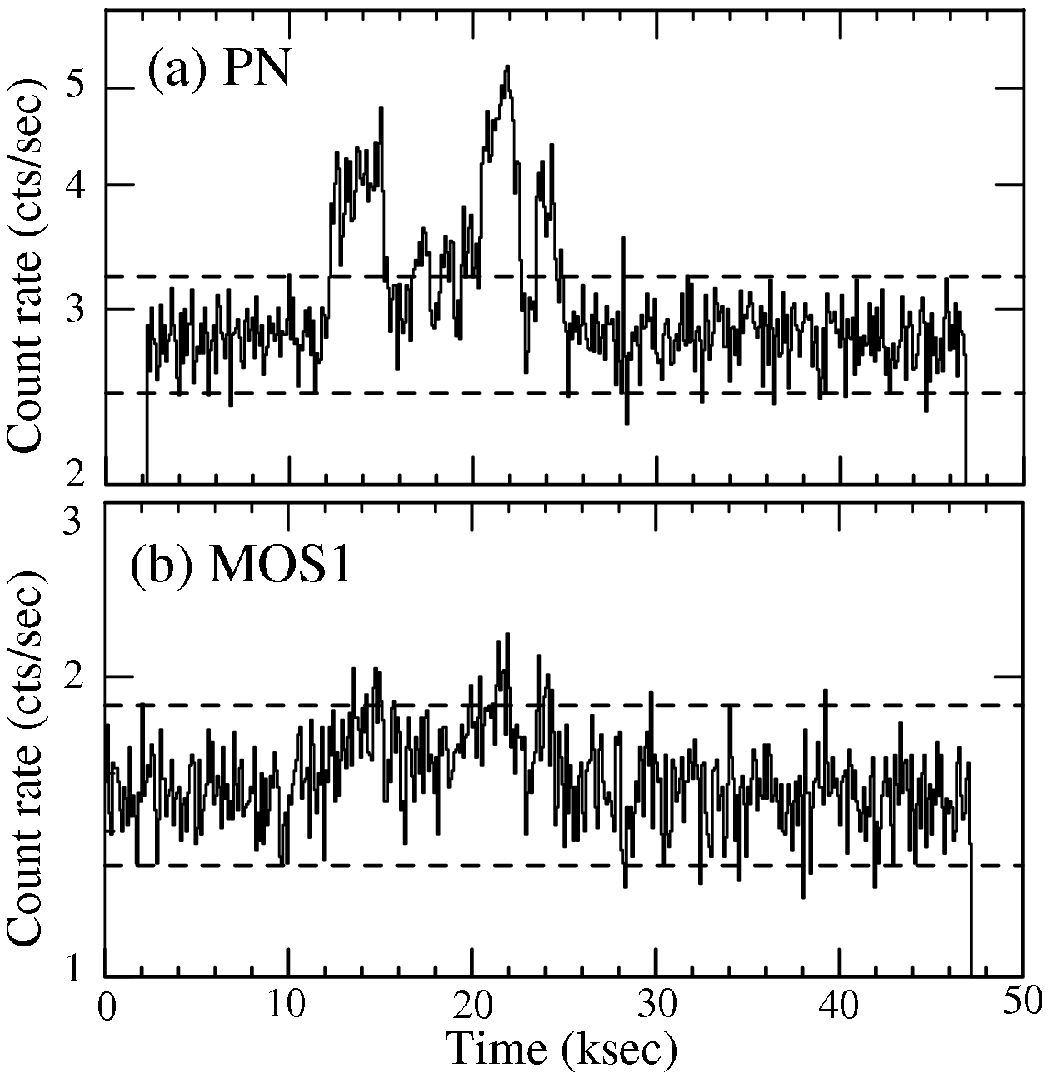}
\caption{
EPIC PN (panel a) and MOS1 (panel b) light curves in 2--7 keV
from the present observation,
obtained by excluding the central ($r < 8'$) region.
The bin width is 100~sec.
The dashed lines indicate $\pm 2 \sigma$ levels from the mean values
in the quiescent periods.
}
\label{fig:lc}
\end{center}
\end{figure}

Soft protons often cause the EPIC background 
to increase suddenly by up to two orders of magnitude.
These sporadic  ``flares'' must be removed carefully
when analyzing extended sources.
We accordingly produced 2--7~keV band light curves of the present data,
excluding a central region~($r\lesssim 8'$) 
where the cluster emission dominates.
This particular energy band was shown by \cite{mpebgd}
to serve as a good measure of the EPIC background,
and is roughly optimized in the present case 
to provide a high signal-to-ratio for proton flares;
below $\sim 2$ keV the cluster signals become significant,
while above $\sim 7$ keV the fluorescent background lines in 
the PN spectra become a nuisance.
The results are shown in Figure~\ref{fig:lc}.
Following \citet{mpebgd} and \citet{itakaphd},
we then discarded those time periods
when the count rates deviate by more than $2~\sigma$
from those in quiescent periods.
After this screening, 
the net exposure time became 31~ksec with PN and 43~ksec with MOS.

Several faint point sources are found in the field of view.
We excluded regions where their signals are significant,
typically within $15''$ of each source.

\subsubsection{Background} \label{subsub:EPIC_bkgd}

Since the cluster emission fills nearly the entire EPIC field of view,
background spectra must be constructed using some other  ``blank sky'' data.
For this purpose,
we utilized the observation of PG~1115+080 for PN~(ObsId=0082340101),
and that of Vik~59 for MOS~(ObsId=0107860501),
because these data sets showed a close similarity 
to the present Centaurus data
in terms of high-energy (11--14~keV for PN and 10--12~keV for MOS) count rates
accumulated over peripheral regions of the fields of view.

After removing  proton flares and  celestial point X-ray sources in the same manner,
we extracted background events from these data sets,
using the same regions on the detectors
as employed to accumulate the on-source data.
The background derived in this way is 
estimated to  be accurate generally to within $\sim 3\%$,
except at some particular spectral regions \citep{mpebgd},
as detailed in \S\ref{subsub:syserr}, where we employ 8\% .
This accuracy is sufficient for our purpose, 
because the cluster emission is much brighter than the background,
e.g., by more than a factor of five,
at least up to $\sim 5'$ from the aimpoint
in energies below 5 keV.
Thus, the background uncertainty therein is 
at most a few percent of the cluster signal.

\subsubsection{Deprojection \label{subsub:depro}}
To discriminate the 1P and 2P conditions,
it is essential to remove the projection effects,
using so-called deprojection procedure.
To create deprojected spectra,
we employed the standard ``onion peeling'' method~\citep[e.g.][]{ikebe04},
assuming that the object is spherically symmetric,
and the plasma is uniformly distributed in each thin shell region.
No model spectra were assumed in our deprojection procedure.
Specifically, at each energy $E$, 
we calculated the deprojected spectrum $S_j(E)$ in the $j$-th shell as
\begin{equation}
S_j (E) = \sum_{k \ge j}^{N} {\cal D}_{j,k} A_k(E)~,
\label{eq:deprojection}
\end{equation}
where $A_k(E)$ is the projected spectrum accumulated over  the $k$-th annulus,
$\{{\cal D}_{j,k}\}$ is a triangular matrix derived by inverting the projection matrix,
and $N$ denotes the outermost annulus.

Below,
we utilize the data within $12'$ (150 kpc) of the X-ray centroid at  
$(12^{\rm h} 48^{\rm m} 49^{\rm s}.2, -41^\circ 18' 44'')$,
which coincides with the NGC~4696 nucleus.
The emission outside $12'$, $S_N (E)$,
was calculated assuming 
that the plasma is isothermal and isoabundance at $R>10'$,
and that  the ICM density profile is expressed by a $\beta$ model
determined jointly with {\it ASCA} and {\it ROSAT},
namely $\beta=0.57$ and the core radius of $7'.3$~\citep{ikebe99}.
These parameters can reproduce  the {\it XMM-Newton} 
images themselves~\citep{itakaphd}.
Actual matrix elements of $\{ {\cal D}_{i,j} \}$
are numerically given in Appendix A.

\subsubsection{Systematic Errors 
\label{subsub:syserr}}
 In order to take into account calibration uncertainties,
conservative systematic errors of 2\% \citep{epicbgd02}
were assigned to the  on-source spectra in energies above 0.6~keV.
Below that energy, 
the systematic error was increased to 5\%
to represent uncertainties in the quantum efficiency
around the oxygen absorption edge~\citep[e.g.][]{mpebgd}.

To further consider systematic uncertainties
involved in the background subtraction process, 
we assigned to the background data 
the systematic errors as given in Table \ref{tbl:syserror}.
The errors are thus taken to be 8\% 
(\S\ref{subsub:EPIC_bkgd}) in those energy bands
where the background is dominated by instrumental fluorescence lines,
because their intensities are known to vary rather 
independently of that of the background continuum.
In energies below $\sim 1$ keV,
the background spectra are significantly contributed
by diffuse soft X-ray background,
of which the brightness varies from sky to sky.
There, we hence employed  the same 8\% systematic error,
although  the background in this energy range is generally 
negligible compared to the bright cluster emission.
In other energy regions,
we employed a nominal value of 3\% (\S\ref{subsub:EPIC_bkgd}).

\subsection{RGS Data Reduction \label{sub:RGSana}}
Again using the SAS software,
we processed the RGS data,
and extracted the first and second order spectra
within $r=60''$ (in the cross-dispersion direction) of the center.
The spectral range of 6--23~\AA\ was analyzed,
in order to utilize many atomic lines therein.

The RGS background was derived from blank sky fields
which were made available by the instrument team~\citep{rgsbgd}.
Because these authors reported that the rms fluctuation of the RGS background
is not more than 30\% over the full spectral range,
we assigned a systematic error of 30\% to the RGS background.
In reality, the emission from the Centaurus cluster exceeds the background
by more than an order of magnitude over the interested spectral range;
therefore, the background uncertainties have little effects 
on the results.

\section{DATA ANALYSIS AND RESULTS}

\subsection{EPIC Data Analysis \label{sub:EPIC_ana}}

\subsubsection{EPIC Spectra}
\label{subsub:EPIC_spec}

Figure~\ref{fig:rawpspec} shows 
the PN spectra of the Centaurus cluster,
obtained by accumulating the on-source events 
over a series of concentric annular regions 
centered on the NGC~4696 nucleus (\S\ref{subsub:EPIC_process}),
and then  subtracting the background 
described in \S\ref{subsub:EPIC_bkgd}.
The radii defining these annuli are given in the figure caption.
The figure also shows their ratios to those
averaged over an outer region of $5'<r<12'$,
where the cool component is hardly seen.
The spectra from these regions do not bear strong 
atomic lines except the Fe-K  line at $\sim 6.7$ keV,
and their spectral shapes are very similar to one another.
Therefore, these  regions at $r \gtrsim 5'$ must be
filled with an approximately isothermal hot plasma.
The figure also confirms correct background subtraction,
because wrong   subtraction
would make the hardest-band ($\gtrsim 7$ keV) spectra
significantly depend  on the radius.

Toward the center, in contrast,
we can see prominent increases in the line intensities;
the Fe-K line enhancement is caused by the metallicity increase,
while those of Si-K (at rest-frame energies of 1.86 and 2.01 keV), 
S-K (2.46 and 2.62 keV), 
and Fe-L ($\sim 1$ keV) lines are mainly 
due to the emergence of the cool component.

At the very central region ($r< 1'$), 
hereafter called ``cD region'',
Fe-L lines with energies below 1~keV become prominent,
indicating the presence of a still cooler plasma
with a temperature of $\lesssim 1$~keV.
The same inference can be obtained from the drastic increases 
in the H-like to He-like intensity ratios of the Si-K and S-K lines.
We therefore  qualitatively reconfirm
the {\it Chandra} detection of the coolest
($\sim 0.7$ keV in temperature) component at the center~\citep{cencxo}.

Applying the procedure described in \S\ref{subsub:depro}
to the spectra in Figure~\ref{fig:rawpspec},
we derived deprojected spectra from 11 concentric  shells.
Some of them are presented in Figure~\ref{fig:rawdspec},
in comparison with the non-deprojected spectra.
The deprojection obviously degrades the data statistics,
but does not alter the spectral shape very much;
the only eye-catching difference is the reduction in the 
hard X-ray ($>3$ keV) flux of the innermost spectra (denoted [cd]),
caused by the removal of contributions from the
foreground/background hot  emission.

\subsubsection{Single-phase analysis \label{subsub:ana1P}}

As first-cut evaluation of the  annular spectra of Figure~\ref{fig:rawpspec},
we conducted a conventional single-temperature analysis. 
This is meant to approximate  1P conditions, but only crudely, 
because three effects, coupled with the temperature and abundance gradients,
would make  each spectrum deviate from an exact isothermality;
the most dominant one is the  projection effects, 
while the less important ones are  
the finite angular resolution of the X-ray telescope
which mixes up signals from adjacent shells,
and  the finite thickness of each annulus.
Thus, the  single-temperature representation
should be regarded as a crude approximation here.

We  fitted the three projected EPIC spectra from each annulus
with a single-temperature APEC model~\citep{apec} 
with photoelectric absorption,
simultaneously but allowing the model normalization 
to differ among the three detectors.
The abundances of O, Ne, Mg, Si, S, Ar, Ca, Fe, and Ni were all left free, 
while those of C and N were fixed at the solar values by \citet{abun}.
The redshift was fixed at 0.0104,
and the hydrogen column density  at the Galactic-line-ofsight  value,
$8.1\times 10^{20}$~cm$^{-2}$,
because \cite{matsushita07aap} found the absorption 
to be grossly consistent with it.
Since some systematic residuals were observed around the Fe-K line region,
we incorporated small gain correction factors, 
+0.3\% to PN, +0.6\% to MOS1, and +0.3\% to MOS2,
all  within the reported calibration uncertainties.
 These improved the fit chi-squared,
without significantly  affecting the best-fit parameters.

 As expected from Figure~\ref{fig:rawpspec},
fits to the cD region spectra left large residuals around the Fe-L energies.
We therefore added, in these cases,  another cool APEC component
with a temperature  (left free) of $\sim 0.7$ keV,
of which the abundances are  tied to that of the major component.
As a result, the fits were much improved,
in agreement with the report by  \cite{cencxo}.
This modeling should be distinguished from the 2P approach to be examined later.

The 1P model  (plus the 0.7 keV component)
reproduced the spectra  moderately well,
although the fits are not necessarily acceptable.
Figure~\ref{fig:1prad} (filed circles) shows 
radial profiles of the obtained parameters.
Thus, the plasma temperature approaches 
a constant value of $\sim$~3.8~keV toward outer regions, 
while drops inside $5'$,
in agreement with the inference from Figure~\ref{fig:rawpspec}b
and with the previous results
\citep[e.g.][]{fabian94asca,fuka94,ikebe99,furusho01}.
As mentioned above, the $\sim 0.7$ keV plasma component 
was needed at $r < 1'$.

In Figure~\ref{fig:1prad},
the abundances of Fe and Si  are seen
 to increase toward the center as first revealed with {\it ASCA}~\citep{fuka94},
while that of oxygen is spatially rather constant.
These results are consistent with the single-temperature 
analysis results from \cite{matsushita07aap}.
Although \cite{cencxo} reported a {\it Chandra} detection 
of a  metallicity decrease within $r\sim 45''$
(down to $\sim 0.4$~solar at $r<10''$),
such an effect, if any, is less significant in the EPIC data.

We performed the same model fitting to the deprojected spectra.
Since the spectra are now free from the projection effects,
the single-temperature modeling is  equivalent to 1P conditions,
assuming that each deprojected shell is thin enough
and  the signal mixing due to the telescope response is negligible.
This analysis yielded the data points represented 
in Figure~\ref{fig:1prad} by open diamonds.
Thus, the overall results are very similar to 
those obtained with the  annular spectra.
The fits became generally better, 
due to the reduced statistics of the deprojected spectra,
and to the removal of fore- and background contributions.
The latter also makes the deprojected temperatures systematically
lower (typically by 0.3--0.5 keV) than the non-deprojected values.

\subsubsection{Two-phase analysis \label{subsub:ana2P}}
As an alternative approach,
we fitted the  EPIC spectra  by a 2P model,
i.e., a sum of two APEC components with different temperatures,
plus the $\sim 0.7$~keV component at $<1'$.
For each annulus,
the cool component temperature $T_{\rm c}$ was left free,
while the abundance of each element was assumed
to be the same among the  components.
Like in the 1P analysis~(\S\ref{subsub:ana1P}),
the  C and N abundances were fixed;
so were the redshift and the photoelectric absorption.
When we first let the hot component temperature $T_{\rm h}$ float as well,
it became rather unconstrained in some outer regions,
and scattered over a range of  $\sim 3$ to $\sim 7$ keV.
This is due to coupling with $T_{\rm c}$ \citep{matsushita07aap}.
However, the value was  generally consistent, within rather large errors,
with 3.8--4.0 keV obtained in past observations.
We hence fixed it at $T_{\rm h} = 3.8$~keV
after the {\it ASCA} results~\citep{ikebe99}
and our 1P fits~(Figure~\ref{fig:1prad}).

By this 2P model,
both the non-deprojected and deprojected spectra
were reproduced reasonably well in all regions.
The obtained parameters,  shown in Figure~\ref{fig:2prad},
are generally  consistent with the two-temperature
results by \cite{matsushita07aap} within some differences in the modeling.
Thus, the fit goodness is comparable to,
or slightly better than,
those obtained with the 1P analysis.
The derived abundance profiles are 
essentially the same as  the 1P results,
and the radial profile of Fe  (Fig.~\ref{fig:2prad}c) agrees
very well with the {\it ASCA} 2P results \citep{ikebe99}.
The temperature differences between the projected and non-deprojected 
spectra became smaller than in the 1P analysis,
and often insignificant within errors.
This is consistent with a 2P condition,
in which an annulus and the  corresponding 
deprojected shell are expected to 
share nearly the same two temperatures.

Figure~\ref{fig:vfill}a shows radial profiles 
of the emission measures (per unit volume) of the hot and cool components,
denoted $Q_{\rm h}$ and $Q_{\rm c}$, respectively.
Thus, the cool component is required  within $6'$,
where the 1P temperature decreases significantly
in our 1P result (Figure~\ref{fig:1prad}).
The  derived  $T_{\rm c}$ stays at $\sim 2$ keV over $2'-6'$
(though rather poorly determined outside $3'$; Figure~\ref{fig:2prad}a),
while it  decreases down to $T_{\rm c} \sim 1.6$ keV at the center.
The latter agrees with  the central 1P temperature 
from our 1P fits (Figure~\ref{fig:1prad}) and from {\it ROSAT}~\citep{cenrosat}.
At the cD region, 
the third 0.7 keV component is again required.
In contrast, neither the ring-sorted spectra nor the deprojected 
thin-shell ones required the cool component in regions outside $6'$.
Accordingly, Figure~\ref{fig:2prad}  outside $6'$
simply reproduces the 1P results.

Incidentally,  the 2P analyses of the {\it ASCA} data 
yielded somewhat lower values of $T_{\rm c}$;
$\sim 1$ keV \citep{fuka94}, 
or $1.4\pm 0.2$~keV~\citep{ikebe99}.
This is partially  because {\it ASCA} was not able to 
separately detect the 0.7~keV component.
The remaining difference may be due to
the different  plasma codes \citep{matsushita07aap},
the MEKAL and APEC codes used 
in the  {\it ASCA} study and the present analysis, respectively.


To characterize the obtained 2P solution,
let us  introduce, after \citet{ikebe99},
the volume filling factor $\eta_{\rm c}$ of the cool phase.
Using the emission measures, 
this quantity is  defined as
\begin{equation}
Q_{\rm c} = \xi n_{\rm c}^2 \eta_{\rm c}, \quad Q_{\rm h} = \xi n_{\rm h}^2 (1-\eta_{\rm c})~,
\end{equation}
 where $n_{\rm h}$ and $n_{\rm c}$ are the densities
of the hot and cool phases, respectively,
while $\xi$ is a factor of order unity reflecting the metallicity.
Assuming a pressure balance
between the two phases~\citep{fuka94}, namely,
\begin{equation}
n_{\rm c}T_{\rm c} = n_{\rm h}T_{\rm h},
\label{eq:pressure_balance}
\end{equation}
we can calculate $\eta_{\rm c}$ as
\begin{equation}
\eta_{\rm c} = \left[ 1+ \left(\frac{T_{\rm h}}{T_{\rm c}}\right)^2 
  \left(\frac{Q_{\rm h}}{Q_{\rm c}}\right) \right]^{-1}.
\label{eq:filling_factor}
\end{equation}

 Figure~\ref{fig:vfill}b shows the radial profile of $\eta_{\rm c}$,
thus calculated using eq.(\ref{eq:filling_factor});
in deriving this result,
we fixed $T_{\rm c}$ at 2.0 keV in shell regions outside $R=1'$.
The figure reconfirms the {\it ASCA} results \citep{fuka94,ikebe99},
that the hot phase dominates the volume
down to a radius rather close ($R\sim 1'.5$ or 18~kpc) to the center.
The filling factor implied by the {\it XMM-Newton} data is
higher by a factor of $\sim 3$ than the {\it ASCA} determination;
as can be understood via eq.(\ref{eq:filling_factor}),
this is due partially  to the higher value of $T_{\rm c}$.

\subsection{Analysis of Thick-Shell Spectra 
\label{sub:thick_shell}}

As we have seen in \S\ref{sub:EPIC_ana},
the two descriptions, 1P and 2P,
are consistent with each other in the sense
that both find essentially the same set of three plasma temperatures,
$\sim 3.8$ keV, 1.7--2.0 keV, and $\sim 0.7$ keV.
Outside $\sim 6'$ ($\sim 75$ kpc),
the ICM can be approximated as isothermal at $\sim 3.8$~keV.
Within $\sim 6'$ toward the cluster core, 
the ICM becomes gradually cooler down to $\sim 2$ keV 
(if employing the 1P description),
or the $\sim 2$ keV plasma becomes intermixed with the hot 
3.8 keV component with a progressively larger $\eta_{\rm c}$
(if employing the 2P description).
In the cD region, 
both scenarios require the additional 0.7~keV plasma.
Up to this stage, 
we are  unable to tell which of the two approaches
is more favored by the data.
Although the 2P fit to the deprojected spectra often gives
a slightly lower reduced chi-squared than the 1P fit
 (Figure~\ref{fig:1prad}b vs. Figure~\ref{fig:2prad}b),
the difference is not necessarily statistically significant,
and could be due to the finite temperature gradient within each shell.

\subsubsection{Construction of thick-shell spectra
\label{subsub:thick_shell_spec}}

In order to better distinguish the 1P and 2P modelings under higher statistics, 
we took the deprojected spectra
from  consecutive five thin shells covering altogether  $R=1'-5'$
([e] though [i] in Figure \ref{fig:rawpspec}),
which have 1P temperatures of 2.03, 2.48, 2.65, 3.10, and 3.30 keV
respectively (\S\ref{subsub:ana1P}).
We then summed them into a single spectrum
which now represents a thicker 3-dimensional shell in the cool core region;
this is hereafter called ``Shell C''.
The cD region, together with the ``plume" structure,
is excluded to avoid complexity introduced by the  0.7~keV component.
Since the statistical errors associated with the adjacent 
thin-shell spectra are not mutually independent,
we estimated the errors to be assigned to the thick-shell spectra
by properly considering  error propagation 
utilizing eq.(\ref{eq:deprojection})\citep{itakaphd}.

The Shell~C spectra, one for each detector,
should contain the 3.8 keV component
if the 2P scenario is correct.
If instead the 1P picture is more appropriate,
the Shell~C spectra would rule out contributions 
from such a high temperature component, 
because the 1P temperature changes 
across Shell~C from 2.0 keV to 3.3 keV 
but not higher (Figure \ref{fig:1prad}).
We expect that the improved data statistics,
achieved by the spectral summation,
allow us to distinguish these two cases.

In addition to the improved statistics,
analyzing a thick shell has two more advantages:
the signal mixing caused by the finite telescope resolution is reduced,
and the errors due to spectral changes
within individual thin shells, if any, nearly cancel out.
As an instructive exercise of the latter effect,
let us assume, for example, the 3rd and 4th annuli constituting Shell~C  
to have a difference by $\delta W$
in their Fe-K line  equivalent widths,
and examine what would happen 
if we artificially assigned the two annuli with a same value $\bar{W}$
that is the emission-measure-weighted average between the two shells.
This exaggerates the procedure of averaging out 
a radial spectral change over a finite shell thickness.
The errors caused by the averaging are expected to propagate
through deprojection into the 1st through 4th thin shells.
However,  as shown in Appendix A,
a mathematical estimate using actual matrix elements 
of eq.(\ref{eq:deprojection}) confirms
that the errors in the four thin shells roughly cancel out,
and hence the final thick Shell-C spectrum suffers 
in the present case no larger errors in the line equivalent width
than $\sim 0.01 \delta W$.
Since the same estimate applies to a temperature gradient,
which is at most $\sim 25\%$  in each thin shell (Figure \ref{fig:1prad}) over $1'-5'$,
the residual temperature errors in the Shell C spectrum are 
estimated to be at most $\sim 0.3\%$, and hence negligible.

\subsubsection{Single-phase modeling
\label{subsub:thick_shell_1P}}

Let us examine the Shell~C spectra against the 1P view.
We hence took the best-fit single-temperature 
APEC model for each of the 5 constituent thin shells,
and convolved it with a response of each detector
that is  weighted by the surface brightness 
of the projection of the relevant thin shell.
In the same way as the actual thick-shell data,
we then summed up the 5 simulated thin-shell spectra 
of each detector into a single one,
to be called ``synthetic 1P spectrum''.

Figure \ref{fig:regCcmpr}a directly compares these synthetic 1P spectra 
with the actual Shell~C spectra of the corresponding detectors.
Thus, we observe a moderately good agreement,
but the actual Shell~C spectra exhibit some excess 
above the synthetic 1P model toward higher energies.
In fact, their difference, 
in terms of chi-squared summed over the three detectors 
(PN, MOS1 and MOS2), 
becomes $\chi^2/\nu=739/378$ 
as summarized in Table~\ref{tbl:chisquares}.

The data vs. model comparison in Figure \ref{fig:regCcmpr}a 
incorporates no adjustment,
since the synthetic 1P spectra are uniquely specified by the best-fit models 
describing the 5 constituent thin shells (\S\ref{subsub:ana1P}).
However, the agreement might be improved  
by adjusting the model parameters.
We therefore allowed to vary freely 
the normalizations of the five APEC components
which constitute the synthetic 1P spectrum.
Then, the agreement (now to be called ``a fit'') 
was improved to $\chi^2/\nu=724/373$ (Table~\ref{tbl:chisquares});
the normalizations of the five components changed
by a factor of 1.65, 0.25, 0.84, 1.10, and 1.26 
in the increasing order of the temperature,
while the summed normalization remained
unchanged within $\sim 2\%$.
Thus, the coolest and hottest of 
the five components tend to be enhanced,
implying a  2P-like condition.
To retain a stable fit convergence,
we did not attempt to let the 5 temperatures float.

\subsubsection{Two-phase modeling
\label{subsub:thick_shell_2P}}

In order to next examine the 2P approach, 
we fitted the actual Shell~C spectra
with a sum of two APEC models,
exactly in the same manner as in \S\ref{subsub:ana2P};
the two temperatures and the two normalizations were left free, 
as were the abundances of
O, Ne, Mg, Si, S, Ar, Ca, Fe, and Ni
which are assumed to be common to the two components.
The relative normalizations among the 
3 detectors were left free as before.
As a result,  the degree of freedom decreased
by 15 compared to the synthetic 1P modeling.
The two components were convolved with the same response, 
weighted by the X-ray surface brightness.

As shown in Figure~\ref{fig:regCcmpr}b and Table~\ref{tbl:chisquares},
the obtained fit ($\chi^2/\nu = 688/363$)
is significantly better than that with the synthetic 1P model,
even considering the reduced degree of freedom.
In fact, the high energy excess, observed in Figure~\ref{fig:regCcmpr}a,
has been reduced in Figure~\ref{fig:regCcmpr}b.
The two temperatures have been obtained as
$T_{\rm h} = 4.02^{+0.41}_{-0.33}$~keV
and $T_{\rm c} = 2.06^{+0.07}_{-0.10}$~keV.
While these values agree  well with the 2P results 
from the individual thin shells (Figure~\ref{fig:2prad}),
the obtained $T_{\rm h}$ is significantly higher 
than the temperature range 
(up to 3.30 keV, with a typical error of $\pm 0.15$ keV)
involved in the 1P description of Shell~C (Figure~\ref{fig:1prad}).
Although the error ranges associated with the derived $T_{\rm h} $ and $T_{\rm c}$
are somewhat (typically by $\sim40\%$) under-estimated 
due to rather large values of the reduced chi-squared ($\sim 1.9$),
the above inference remains valid
even if the error ranges are enlarged by $\sim40\%$.

Just for consistency,
we applied in the same way the 2P fit to the synthetic 1P spectra,
to which Poissonian noise was added.
This yielded $T_{\rm c} = 2.3$ keV and $T_{\rm h} = 3.3$ keV,
although the associated errors are rather difficult to estimate.
These two temperatures are both contained within the 1P 
temperature distribution across Shell~C (Figure~\ref{fig:1prad}), 
from 2.03 keV to 3.30 keV,
that we employed in constructing the synthetic 1P spectra.
This difference between the actual Shell~C spectrum
and the sythetic 1P spectrum suggests
that the hot 3.8 keV component was  already present in the 5 thin-shell spectra,
and has become  more significant by the spectral summation.

In the above 2P fit to the actual Shell~C spectra,
the two components were treated to have the same abundances,
and were  convolved with the same response.
However, the true 2P condition would be somewhat different,
since the cool component,
more weighted toward the metal-enriched inner regions than the hot one,
must have higher abundances when averaged over Shell~C. 
To better express this condition,
we returned to the original best-fit 2P models to the five thin-shell spectra,
convolved them with projected responses of the respective thin shells,
and summed the results over the 5 shells.
These simulated data,
involving two fixed temperatures (2.0 and 3.8 keV)
and properly considering the radial abundance changes,
are to be called ``synthetic 2P spectra'',
because they are constructed exactly in the same way as their 1P counterparts.
Although these synthetic 2P spectra 
do not involve any adjustable free parameter,
they have reproduce the Shell~C spectra with $\chi^2/\nu=678/378$,
even better than the above 2P fit,
and also than the synthetic 1P modeling.

\subsubsection{Error renormaliztion}
\label{subsub:renormalize}

From the statistical viewpoint,
even the synthetic 2P simulation obtained in \S~\ref{subsub:thick_shell_2P}
does not give an acceptable reproduction of the data (Table~\ref{tbl:chisquares}).
This hampers correct error evaluations,
and makes it difficult to quantitatively compare different modelings.
In Figure~\ref{fig:regCcmpr},
fit residuals  are observed around strong emission lines,
such as Si-K, S-K, and Ar-K.
Therefore, the  inadequate fit goodness is likely to be caused mainly 
by subtle calibration inaccuracy in reproducing the detailed line profiles,
rather than model inappropriateness.
Under the higher data statistics,  
the calibration uncertainties presumably 
exceeded the systematic errors  of 2\%
already assigned in \S\ref{subsub:syserr}.
Although our data analysis utilized rather old SAS version (5.4.1),
a reanalysis of a few representative spectra from the central region,
using an updated version 7.0.0 SAS software (processing and responses),
showed rather insignificant improvements of these problems.
Furthermore, the fitting results derived with the two versions agreed within errors.
Accordingly, we retain the version 5.4.1 analysis, 
and assigned 5\% of the source counts 
as a new systematic errors to the whole energy range,
and repeated the overall analyses.
This procedure is meant to make the fit acceptable 
when using the model that is best preferred by the data,
and to make securer the distinction among different models.

By this error renormalization,
the values of $\chi^2$ have reduced appreciably,
while the radial profiles of various 
plasma parameters remained unchanged.
Specifically, the comparison of the Shell~C spectra 
with the remade synthetic spectra gives
$\chi^2=418$  and 381 for 1P and 2P, respectively, both with $\nu=378$
(Table \ref{tbl:chisquares}).
The latter is statistically acceptable,
and gives a chi-squared which is smaller by 30 than the former.
Thus,  the  2P modeling of the Shell~C spectra is 
statistically superior to the 1P one.

\subsubsection{Thick Shell~P spectra}

In a similar way,
we derived EPIC spectra from another thick shell, called Shell~P, 
which sums up three consecutive thin shells covering $R=3'-6'$.
This Shell~P is peripheral to the central cool region.
The summed Shell~P spectra were then compared with 
the corresponding synthetic 1P and 2P spectra,
constructed in the same manner as before
from the best-fit solutions to the 3 constituent thin shells.
The comparison of the actual spectra 
with the synthetic 1P and 2P spectra gave  
$\chi^2=511$ and 496
(without the error renormalization), respectively, both with $\nu=378$.
The difference in chi-squared,
though smaller than in  the Shell~C case, amounts to 15,
and remains $\sim 11$ even when the systematic error is 
renormalized as in \S\ref{subsub:renormalize}.

We also fitted the Shell~P spectra with the 2P model,
in the same way as for the Shell~C case 
but fixing $T_{\rm h}$ at 3.8~keV
as we did in \S\ref{subsub:ana2P}.
We then obtained $T_{\rm c} = 2.10^{+0.38}_{-0.29}$~keV,
together with $\chi^2/\nu = 494/364$.
This value of $T_{\rm c}$ again agrees with those obtained by the 
2P fits to the constituent thin-shell spectra (Figure~\ref{fig:2prad}),
while it is significantly below the distribution 
of the 1P temperature across Shell P,
3.0--3.5~keV (Figure~\ref{fig:1prad}).
Therefore, the cool ($\sim$~2.0~keV) component is inferred 
to be present even in this relatively outer region.
As a consistency check,
we fitted  the synthetic 1P spectra in Shell P 
with the same 2P model,
and obtained  $T_{\rm c} = 3.2$~keV
which agrees with the input 1P temperature range. 

\subsubsection{Possible artifacts}
\label{subsub:artifacts}
We have so far found 
that the two sets of thick-shell spectra
both prefer the 2P view to the 1P scenario.
However, the results could be subject to various artifacts.

An immediate suspect is
that the appearance of the hot component in Shell~C is 
an artifact caused by insufficient background subtraction.
However, this would not explain the presence
of the cool component in Shell P.
Furthermore, the  constant spectral shape in  $r > 5'$ 
(\S\ref{subsub:EPIC_spec}; Figure~\ref{fig:rawpspec})
argues against this possibility.
For a more quantitative examination,
we remade the deprojected spectra
with the background artificially under-subtracted, 
by 3\% which is comparable to the background systematic errors
estimated in \S\ref{sub:EPIC}.
In fact, this little affected the results obtained so far:
a comparison of the remade Shell~C spectra
with the 2P synthetic spectra (\S\ref{subsub:thick_shell_2P})
gives nearly the same $\chi^2$ as before,
changing only by $\sim$~2.
Thus, our results are robust
against the background uncertainty.

Another concern is possible deviations 
from the so-far assumed spherically symmetry.
Indeed, the ICM temperature of the Centaurus cluster exhibits 
significant non-axi-symmetric distributions~
\citep{furusho01,molendi02,cencxo,cencxo2}.
Then, even if the ICM were in a 1T condition,
Shell C would sample a wider range of temperatures,
to mimic a 2T condition.
However, as far as we consider an annular region of $4'< r <8'$
that is crucial to this issue,
the ICM temperature (approximated as 1T) is azimuthally constant 
typically within $\pm 15\%$ \citep{furusho01}.
To examine this issue using the present data,
we derived  a temperature map, presented in Figure~\ref{fig:Tmap},
by analyzing the projected EPIC spectra
in four sectors by a single-temperature model.
It reconfirms a mild temperature decrease
from northwest to southeast directions,
with an amplitude comparable to those measured previously.
According to a simple estimation,
a temperature variation of this amount is insufficient, by at least a factor of two,
to explain the difference between the two values
of $T_{\rm h} = 4.02^{+0.41}_{-0.33}$~keV and $T_{\rm h} = 3.03$ keV,
obtained via the 2T fit to the  Shell~C spectrum
and its synthetic 1T data, respectively.
Thus,  the preference for the 2T view remains intact.

The finite angular resolution of the {\it XMM-Newton} X-ray telescope
could also blur the Shell~C boundary,
and cause contaminations by signals
from the outer hotter region.
However, Shell~C is much thicker ($4'$) than the blurring width ($\sim 15"$),
and the  surface brightness decreases outwards.
As a result, the contamination of the outer hotter emission
into Shell~C is estimated to be at most $\sim 2\%$,
which is within the renormalized systematic uncertainty of 5\%
introduced in \S~\ref{subsub:renormalize}.

In the deprojection process (\S~\ref{subsub:depro}),
we assumed  the region outside $12'$ (150 kpc) 
to have a constant spectrum 
that is identical to that  measured just inside it.
This is based on the fact that  the azimuthally-averaged ICM temperature of this cluster,
measured with the {\it ASCA} GIS
utilizing its wide field of view and low background,
is essentially constant at $\sim 3.9$ keV 
from $r=14'$ up to $r=40'$  \citep[0.5 Mpc; ][]{ikebe99}.
Of course, deviations from this assumption 
would in principle affect, via eq.(\ref{eq:deprojection}),
the deprojected spectrum in every shell.
However, our numerical estimate in Appendix A indicates
that the emission from the $R>12'$ region contributes
only $\sim 1\%$ to the Shell C spectrum,
and hence any realistic error associated to it is considered negligible.
Furthermore, the ICM temperature has been confirmed in many clusters
to decrease beyond  $\sim 0.2$ times the virial radius
\citep[e,g.,][]{outerkT96, outerkT02, suzaku1060, outerkT08},
or $r \gtrsim 0.5$ Mpc in the present case.
Then, we are likely to be overestimating the outermost ICM temperature,
and hence over-subtracting the hot (3.8 keV)  emission 
as foreground and background contributions.
This makes our argument simply more conservative.

A more difficult issue
is  asymmetry along the line of sight.
If the object has a prolate shape in the depth direction,
the deprojected shell~C spectra would sample 
effectively outer (hence hotter) regions,
and would involve the 3.8 keV component
even if it is in a 1P condition.
Similarly, an oblate 1P condition would
make Shell~P sample the cool emission.
However, neither case can
explain the two thick shells simultaneously.

Finally in \S~\ref{subsub:thick_shell_spec},
possible spectral changes within each thin shell was already shown 
to become negligible through the construction of thick shells.
As an overall confirmation to address this issue,
in Appendix B we have numerically simulated a 1P cluster,
and analyzed the fake data in the same manner as for the actual data.
Then, the simulated Shell C spectra have been reproduced by the 1P view,
without demanding the 2P formalism.
As a logical contraposition,
the preference of the 2P modeling by the actual data demands
the actual object not to be in the 1P condition.

From  these examinations, we conclude
that the data favor the presence of the
hot $\sim 3.8$~keV component  in the 3-dimensional core region ($R=1'-5'$),
and the cool $\sim~2$ keV component in the peripheral region ($R=3'-6'$).
In other word,  the 2P picture is considered to
describes the actual physical state better than the 1P view.

\subsection{Excess Absorption
\label{sub:excess}}
At the center of the Centaurus cluster,
some previous works reported the presence of excess 
X-ray absorption, 
up to a few times $\sim 10^{21}$ cm$^{-1}$
(depending on the modeling),
above the Galactic  column
\citep{cenrosat,fabian94asca,cencxo};
this was taken as evidence for the CF.
However, the effect was not
confirmed in other works \citep{ikebe99}.
Likewise, we have here experienced no problems in our 
spectral analysis  with the absorption fixed at the Galactic value.

Generally, 
a somewhat higher absorption is often derived
when a two-  (or multi-) temperature emission 
is fitted with a 1P model  \citep[e.g.,][]{matsushita07aap}.
However, some of the previously reported 
excess  column densities in the Centaurus cluster
considerably exceed such modeling uncertainties.

We examined this issue using the non-deprojected and deprojected spectra
from regions within $10''$ and $30''$ of the center, respectively.
Specifically, we refitted these EPIC data 
using the 2P plus 0.7 keV  model,
applying an intrinsic absorption factor to both components
in addition to that due to the Galactic column.
(The hot component can be neglect here.)
As a result, the fits were little improved in either case.
The obtained {\it excess} column density in the rest frame 
is $3.2^{+2.0}_{-1.8}\times 10^{20}$~cm$^{-2}$
for the non-deprojected spectra within $10''$,
and $0.5^{+2.3}_{-0.5}\times 10^{20}$~cm$^{-2}$
for the deprojected spectra within $30''$.
These are significantly lower than the previously claimed values,
$\gtrsim 1\times 10^{21}$~cm$^{-2}$, 
and the case of the deprojected spectrum is consistent
with no excess absorption.
Then, what is responsible for this discrepancy?

Regardless of the modeling (1P or 2P),
the {\it XMM-Newton} data reveal
that oxygen in the central region
is deficient relative to iron and silicon
(\S\ref{subsub:ana1P}, \S\ref{subsub:ana2P}).
Nevertheless, the previous works usually assumed  
metals in the ICM to obey the solar abundance ratios,
simply because of inadequate energy resolution.
This presumably caused data deficits around the 
oxygen line ($\sim 0.65$ keV) relative to the model predictions,
and forced previous investigators
to introduce an artificial excess absorption
in an attempt to suppress the over-predicted O-K line flux in their models.
Actually, if we constrain the ICM to follow the solar abundance ratios,
the present data clearly require an excess absorption
by $\sim 1.4\times 10^{21}$~cm$^{-2}$,
although the fit becomes worse (with $\chi^2/\nu=1.38$)
compared to the case of the Galactic absorption 
and free  abundance ratios ($\chi^2/\nu=1.13$).
Thus,  the reported excess absorption 
in the Centaurus cluster is considered an artifact,
caused mainly by the non-solar 
abundance ratios around the cD galaxy.

\subsection{RGS Results
\label{sub:RGSresult}} 

Because the surface brightness of the Centaurus cluster is
strongly peaked at the center,
we can utilized the RGS data to better constrain 
the temperature structure of the ICM in the cD region, 
although the available spatial information is 
limited to 1-dimensional projections.
We  extracted the first and second order spectra from RGS1 and RGS2,
over a strip of $2'$ in width centered on the NGC~4696 nucleus.
The cross-dispersion direction is at a position angle of $\sim 20^\circ$,
so that the data accumulation strip partially covers the plume.
The background  was subtracted
as described in \S\ref{sub:RGSana}.

Figure~\ref{fig:rgs}a shows the obtained RGS spectrum,
which combines the 1st and 2nd order spectra from the two RGS units.
Although the energy resolution is somewhat degraded
due to the finite angular extent of the source,
we clearly observe many atomic emission lines.
In particular, the spectrum bears a strong K-line 
at $\sim 19$~\AA~ from H-like oxygen ions (OVIII),
but lacks those from He-like ones (OVII)
which would be emitted strongly at about 22~\AA~
by plasmas with a temperature lower than 0.3~keV.
This indicates that the cluster core region 
is devoid of such very cool plasmas,
contrary to the prediction by the CF hypothesis.

Since we already know
that at least two plasma temperatures (0.7 keV and 1.7--2.0 keV) 
are required to reproduce the emission from the cD region,
we fitted the RGS spectra with a simple 2P plasma emission model;
we hereafter call this particular modeling ``quasi-2P fit'',
because it indeed invokes two APEC components 
but  the employed temperatures are different from those composing  
the 2P model employed in \S\ref{subsub:ana2P} and  \S\ref{sub:thick_shell}.
The four spectra (1st- and 2nd-order spectra from RGS1 and RGS2)
were prepared separately, and fitted jointly.
We neglected for the moment the hot (3.8 keV) component,
because its high temperature and large angular extent
make low-energy lines rather weak,
and hence its parameters are difficult to constrain with the RGS data.
The abundances of O, Ne, Mg, Si, Fe, and Ni were left free,
but constrained to be common between the two components.
After the EPIC results \citep{itakaphd},
we fixed the S, Ar, and Ca abundances 
at 1.7, 1.5, and 2.4~solar, respectively, 
because K-lines of these elements are outside the RGS wavelength range.
The absorption is  assumed to be Galactic,
as confirmed in \S\ref{sub:excess}.
To deal with the spatial extent,
we blurred each spectral component
with a Gaussian of a free width $\sigma$.

The quasi-2P fit to the RGS spectra 
is shown in panels (b) and (c) of Figure~\ref{fig:rgs},
and the obtained parameters are summarized in Table~\ref{tab:rgsfit}.
Thus, the overall emission-line features 
have been reproduced to a reasonable extent.
Although the fit is not yet fully acceptable,
this is mainly due to some unessential factors,
such as our simple spectral blurring, 
the projection effects, 
and calibration uncertainties.

The obtained two temperatures, 0.8~keV and 1.7~keV, 
agree well with the lowest two characteristic values 
measured with the EPIC from the cD region.
The Gaussian widths of the 0.8~keV and 1.7~keV components,
$\sim 20''$ and $\sim 65''$ respectively,
can be understood as reflecting the angular extent 
of the $\sim 0.7$ keV and $\sim 1.7$ keV component 
detected with the EPIC.
The RGS-determined abundances of Mg, Si, and Fe are lower 
by a factor of $\sim 2$ than the values obtained with the EPIC data.
The discrepancy is at least partially relaxed 
by considering the hot~(3.8~keV) component,
which does not emit prominent lines,
but is estimated to contribute to the continuum up to 30\%
and reduce equivalent widths of the lines by that amount.
The remaining abundance discrepancy
could be due to the abundance drop
at the very center  \citep{cencxo}.
In any case,  the  temperatures,
 determined primarily by the line intensity ratios,
are not affected.

The RGS spectra may be examined for 
the presence of  other temperature components,
besides the $\sim 0.8$ keV and 1.7~keV ones constituting the quasi-2P fit.
The RGS spectra were therefore analyzed
with the multi-temperature fit procedure of \citet{tamura03}, 
which is a kind of differential emission measure analysis
like those in \citet{peterson03} and \citet{kaastra04}.
Specifically, we prepared seven plasma emission components
of which the temperatures are given as 
$T_0$, $1.5T_0$, $(1.5)^2T_0$, .., and $(1.5)^6T_0$,
with $T_0$ being the lowest temperature.
We then allowed to vary the normalizations of the seven components,
as well as $T_0$ to which the remaining 6 temperatures scale.
The abundances of O, Ne, Mg, Si, Fe, and Ni were left free,
but constrained to be common among the seven components.
Each spectral component was blurred with a Gaussian,
of which the width is left free but constrained to be less than   
$66''$ as obtained in the previous fit~(Table~\ref{tab:rgsfit}).

As shown in Figure~\ref{fig:rgsdem},
the best fit model obtained from this analysis is physically 
very close to that obtained with the quasi-2P model.
The fit has been improved little 
($\chi^2/\nu = 976/718$; cf. Table~\ref{tab:rgsfit})
by considering extra temperature components,
with only the two components
($\sim 0.8$~keV and 1.7~keV) remaining significant.
Note that these two temperatures are constrained
to have a ratio of 1.5$^{2} = 2.25$, 
while their absolute values are left free.
The 1.1~keV component has turned out to be weak
in contrast to the two adjacent ones,
implying that the significantly detected 
two components are discrete entities
rather than representing a continuous temperature distribution.
The 2.55 keV component is not significant, either. 

The RGS results constrain the 3.8 (=$1.7 \times 1.5^2$) keV 
component to have an emission measure 
which is $< 30\%$ of that of the 1.7 keV component.
Employing eq.(\ref{eq:filling_factor}),
the volume filling factor of this component is 
then estimated as $< [1+(1.7/3.8)^2 / 0.3]^{-1} = 0.6$
(or equivalently $\eta_{\rm c}>0.4$).
This is consistent with Figure~\ref{fig:vfill}.

Figure~\ref{fig:rgsdem} also reconfirms
that any cool emission below 0.7~keV is insignificant.
The upper limits are by more than an order of magnitude 
below the prediction of the isobaric CF model \citep{isobaric},
if the model normalization is adjusted to the data at 1.7 keV.
Therefore, we conclude that the ICM around NGC~4696
is not cooling to temperatures much below $\sim 0.7$ keV.  
Analyzing the RGS data from the present and an additional 110 ks observation, 
\cite{cenrgs} also confirmed 
that the emission measure from the central region
decreases more steeply toward lower temperatures
than in a simple CF picture.

\section{DISCUSSION}

\subsection{Summary of the Obtained Results}

We analyzed the {\it XMM-Newton} EPIC  data 
of the central region ($<12'$ or $<150$ kpc) of the Centaurus cluster,
assuming a spherical symmetry.
The ICM  has been characterized by three representative  temperatures;
 $T_{\rm h} \sim 3.8$ keV, 
$T_{\rm c} = 1.7-2.0$ found within $\sim 6'$ (75~kpc),
and $\sim 0.7$ keV localized within $\sim 1'$ (12~kpc) of the center.
The values of $T_{\rm c}$ and $T_{\rm h}$
are consistent with the {\it ASCA} measurements~\citep{ikebe99},
and satisfy eq.(\ref{eq:TcTh}).
Similarly, the detection of the 0.7 keV component 
reconfirms the {\it Chandra} result \citep{cencxo}.
The RGS data from the core region 
reinforced the presence of these discrete temperatures,
and gave tight upper limits to emission 
with a temperature of $<0.5$ keV.

\subsubsection{2T vs. 1T preference}
\label{subsub:2Tvs1T}

As far as the deprojected thin-shell spectra are individually analyzed,
the data had no preference between the 1P and 2P approaches.
However, the two sets of EPIC spectra
from  3-dimensional thicker shells have revealed 
that the hot component is present 
even in a very inner ($R< 5'$) region,
and the cool component protrudes out 
beyond $R \sim 3'$  (\S\ref{sub:thick_shell}).
The RGS results from the cD region (\S~\ref{sub:RGSresult})
provides an important support to this view.
\cite{itakaphd} found that  {\it XMM-Newton} spectra of the central regions 
of Abell~1795 also prefer the 2P formalism to the 1P modeling.

In principle, the ICM could be in multi-temperature conditions \citep{kaastra04}.
Since a spectrum resulting from such a superposition of a range of temperatures
is rather insensitive to the way of superposing them \citep{DEM76},
an apparently 2P plasma  may well be  in a multi-temperature condition,
and vise versa,
particularly when the energy resolution is not too high.
Indeed,  the measured value of $T_{\rm c}$ appears 
to distribute from 1.7 to 2.0 keV, 
depending on the radius.
Nevertheless, the RGS data  preferred 
the presence of discrete temperatures, 
to a continuous temperature distribution (\S \ref{sub:RGSresult}).
Therefore, a multi-phase modeling of the ICM is
considered inappropriate,
at least the cD region of the Centaurus cluster is concerned.

Based on these results, 
we suggest that the plasmas in the central $\sim 75$ kpc 
of the Centaurus cluster are better envisaged
by invoking a few (two to three) discrete temperature components,
rather than assuming a continuous radial temperature gradient
or a condition of differential emission measure.
These results support and extend our view of 
cD clusters developed with {\it ASCA} (Paper I).
Below, we  adopt the 2P view as our working hypothesis.

When integrated up to $R=12'$ (150~kpc),
the hot component, the cool component, and the 0.7 keV one 
have bolometric luminosities of 
$L_{\rm h}^{\rm bol} = 3 \times 10^{43}$ erg s$^{-1}$,
$L_{\rm c}^{\rm bol}= 1.0 \times 10^{43}$ erg s$^{-1}$,
and $L_{\rm 0.7}^{\rm bol}= 6 \times 10^{41}$ erg s$^{-1}$,
respectively.
According to \cite{ikebe99},
$L_{\rm h}^{\rm bol}$ increases by a factor of 2
if integrated further out to  $R=30'$ (375~kpc).

\subsubsection{ICM abundances}
\label{subsub:abundances}

We have determined the  radial abundance profiles of Fe, Si and O in the ICM, 
assuming that the different plasma components 
share the same abundances at a given radius.
Those of the other elements are much more uncertain \citep{itakaphd}.
As first discovered by \citet{fuka94},
and already reported by \cite{matsushita07aap}
by analyzing  the present {\it XMM-Newton} data,
the Fe and Si abundances have been reconfirmed to increase
significantly toward the center (Figure \ref{fig:2prad}),
and stay rather high even at the very center.
The Si/Fe abundance ratio at the center, 
about 1.3 (Figure~\ref{fig:2prad}),
agrees with the {\it ASCA} measurements~\citep{fuka00}.

In contrast to the behavior of Fe and Si,
the O abundance  is radially more constant,
implying a  decrease in the O/Fe ratio toward the center,
in agreement with \cite{matsushita07aap}.
As argued by various authors
\citep[Paper~I; ][]{mushotzky96,fuka98,fuka00,SNtypes,kawahard, n1550},
this can naturally be understood as a result of 
varying contributions from type Ia and type II suparnovae:
the outer-region ICM is contributed significantly by
oxygen-rich products from  type II supernovae,
while that in the inner region by iron-rich products from
type Ia supernovae that occurred in the cD galaxy.
 
So far,  a fair number of other clusters have been known to exhibit
abundance profiles similar to those of the Centaurus cluster,
with a central increase of iron and silicon,
and a flat distribution of oxygen
\citep[e.g.][]{boehringer01,tamura01b,n1550}.
These include in particular recent  {\it Suzaku} observations of 
the Fornax cluster \citep{matsushita07pasj},
Abell~1060 \citep{suzaku1060},
and AWM~7 \citep{suzakuawm7}.
Therefore, the  relative oxygen deficit in the cluter
core region may be a common phenomenon.

\subsubsection{Evidence against Cooling Flows}
\label{subsub:antiCF}

The present results  argue against the CF scenario
from several independent aspects.
These include; the lack of cooling component  (\S\ref{sub:RGSresult});
the lack of excess absorption (\S\ref{sub:excess});
and the radial changes in the abundance ratios (\S\ref{subsub:abundances})
which rule out large-scale inflows of the ICM.
Below, we discuss the first two issues in some more details.

The CF scenario relates the bolometric luminosity, $ L_{\rm CF}$,
of a ``cooling''  portion of ICM  with the mass deposition rate, 
$\dot{M}$, as~\citep{fabian94}
\begin{equation}
 L_{\rm CF} = \frac{5\dot{M}kT_{\rm h}}{2\mu m_{\rm p}},
\end{equation}
where $\mu$ is the mean molecular weight
and $m_{\rm p}$ is the proton mass.
By substituting the measured value of 
$L^{\rm bol}_{\rm c}+L^{\rm bol}_{0.7}=1.1 \times 10^{43}$ erg s$^{-1}$ 
(\S 5.1) into $ L_{\rm CF}$,
we obtain $\dot{M} = 11\ M_\odot$~yr$^{-1}$.
This in itself is consistent with the previous results 
from {\it ROSAT} \citep{edge92,rosatmdot} and  {\it ASCA} (Paper I).
However, we have found no evidence of emission from
any plasma component with a temperature $< 0.5$~keV,
with upper limits  much tighter 
than the CF-predicted emission measure (Figure~\ref{fig:rgsdem}).
Therefore, the CF interpretation fails to explain the strong 
cool emission of the Centaurus cluster, as well as in other clusters
\citep[Paper~I;][]{tamura01a,kaastra01,peterson01}.
The value of $\dot{M}$ should not b e taken
as an actual mass-deposition rate.

We have shown in \S \ref{sub:excess}
that the previously reported excess absorption,
at the center of the Centaurus cluster,
is an artifact caused by the relative oxygen deficit.
In fact, when the non-solar abundance ratios are properly considered,
the absorption became consistent with being Galactic only,
with any  additional column  at most several times $10^{20}$ cm$^{-2}$. 
As discussed in \S\ref{subsub:abundances},
the relative oxygen depletion in the cluster center
appears to be a common feature.
Then, the  excess absorption, 
claimed previously  in many other clusters \citep[e.g.][]{white91},
will be explained away as well.
Indeed, using the RGS spectra, \cite{peterson03} also confirmed
the lack of excess absorption  from several other clusters.
Given these results, 
there is practically no evidence for the  X-ray absorbing materials
at least in nearby well-studied clusters.

\subsection{The 0.7 keV component}
\label{sub:0.7keV}

Let us briefly consider the nature of the  0.7 keV component,
which is localized to the cD galaxy, NGC~4696.
Similar cool  emission comonents are seen 
around other cD galaxies, including M87 in particular, 
but their interpretation has generally been unsettled 
\citep[e.g.,][]{fabian01a,churazov01,m87cool}.

Judging from the angular extent ($\sim 20$ kpc) 
and the  bolometric luminosity ($6 \times 10^{41}$ erg s$^{-1}$),
this 0.7 keV component could be hot interstellar medium (ISM) 
associated with the cD galaxy.
Its temperature is consistent with 
the stellar velocity dispersion of NGC~4696, 274 km s$^{-1}$.
The plasma emitting this component 
(including both the central concentration and the plume)
has an estimated mass of a few times $10^9~M_\odot$~\citep{cencxo},
which is also consistent with this ISM interpretation.
Although the estimated cooling time of this plasma
is as short as $\sim 10^8$ yr~\citep{cencxo},
the radiative energy loss could be supported by such processes 
in NGC~4696 as the supernova heating (a few times $10^{41}$ erg s$^{-1}$),
and the accumulation of stellar mass loss
which is dynamically ``hot'' on the galaxy scale.

In contrast to the above arguments,
the complex filamentary morphology of this component~\citep{cencxo},
including the plume like elongated tail structure,
cannot easily be reconciled with the hot ISM interpretation,
even considering  interactions of the cD galaxy with the ICM.
Similar problems with the cool gas in M87 (the Virgo cD galaxy)
were pointed out by \cite{m87cool},
who showed its close  association with the synchrotron radio arms.
Furthermore, as exploited in Paper~I,
the hot ISM interpretation could be applied better to the cool (2 keV) component.
Considering these,
the 0.7 keV plasma could alternatively be associated 
with some magnetic structures around the cD galaxy,
which in turn could be a result of past nuclear activity in NGC~4696.
Further discussion of this issue is beyond the scope of the present paper.

\subsection{A Magnetosphere Model
\label{sub:magnetosphere}}

The successful 2P picture (\S\ref{sub:thick_shell}, \S\ref{subsub:2Tvs1T})
and the absence of significant CFs (\S\ref{subsub:antiCF}) in the Centaurus cluster suggest
that  its core region (within $\sim 75$ kpc) hosts a cool plasma phase
which is stably intermixed  with the hot phase.
While this  reconfirms a prediction made in Paper~I,
it poses a series of questions to be answered;
(i) how  such a 2P configuration is realized;
(ii) how the cool phase is heated against the radiative cooling;
(iii) how the heating and cooling be balanced in a thermally stable manner;
(iv) what determines the value of $T_{\rm c}$;
and
(v) what produces the scaling of eq.(\ref{eq:TcTh}).
Below, we examine them one by one.

\subsubsection{Needs for ordered magnetic fields
\label{subsub:magfield}}

An immediate issue associated with the 2P scenario is
(i) how to realize such a configuration.
We may suppose that the two phases are intermixed 
on a typical length scale of $a \sim 10$ kpc,
as indicated by various sharp structures seen in
X-ray images of some clusters \citep{fabian01b,n1404cxo,iizukaphd},
including the  ``cold front'' structure \citep{a3667frontmag}.
Then, employing the classical Spitzer heat conductivity,
\begin{equation}
\kappa =  5\times 10^{-7} T^{5/2} ~~~ {\rm ergs~s^{-1}~cm^{-1}~K^{-1}}~
\label{eq:conduction}
\end{equation}
where  $T$ is the temperature in Kelvin,
the time scale of conductive heat exchange
between the two phases would be very short, as
\begin{equation}
\tau_{\rm cond} \sim 
    6 \times 10^5 \left(\frac{n_{\rm e}}{10^{-3} {\rm cm}^{-3}} \right)
                  \left(\frac{T_{\rm h}}{\rm 4~keV} \right)^{-5/2}
                  \left(\frac{a}{10~\rm{kpc}} \right)^2 ~~~ {\rm yr}
\label{eq:conduction_ts}
\end{equation}
\citep{sarazin}.
The two phases would  quickly become isothermal,
unless the heat conduction between them is  suppressed
~\citep{takahara,asai06}.

The most natural way to suppress the heat conduction is to invoke magnetic fields,
as argued, e.g., by \cite{m87cool} to explain the cool plasma in M87.
In fact, the ICM is known to be generally magnetized to a few $\mu$G,
and the aforementioned sharp X-ray features themselves are indicative of 
ordered strong ($\sim 10~\mu$G) magnetic fields \citep{a3667frontmag}.
In the particular case of the Centaurus cluster,
a magnetic field strength up to $\sim 25~\mu$G is reported
from radio observations \citep{taylor07}.
Although these field strengths are not necessarily  
higher than the equipartition value ($\sim 30~\mu$G),
they are more than sufficient to completely 
suppress the heat transport across field lines,
because the gyration radius of 4 keV thermal electrons in a 1 $\mu$G field,
$\sim 10^8$ cm, is more than 10 orders of magnitude
smaller than their classical mean free path for Coulomb scattering
which also determines eq.(\ref{eq:conduction}).
In contrast, the heat conductivity along field lines may well be 
approximated by the Spitzer value, eq.(\ref{eq:conduction}),
unless the field lines are highly tangled~\citep{conduction}.

Based on solar analogy, 
magnetic field lines around a cD galaxy may be classified
grossly into ``open'' and ``closed'' ones \citep{max97},
corresponding to coronal holes and coronae, respectively.
This view relies only on a general  and simple classification 
of the magnetic field-line topology,
without invoking any particular magnetic configuration.
Then,  open-field regions, 
connected to the outer cluster volume,
must be filled with the hot-phase ICM,
and  kept nearly isothermal 
by the efficient field-aligned heat conduction of eq.(\ref{eq:conduction}).
In contrast, closed-field domains,
connected to the cD galaxy 
and thermally insulated from the surrounding hot phase,
can take independent (possibly lower) temperatures of its own.
In addition, the loop interior plasma is expected to be
metal enriched by type Ia supernovae in the cD galaxy.

From these considerations, we speculate that the plasma,
filling numerous  magnetic loops anchored to the cD galaxy,
is observed as the cool phase of the Centaurus  cluster (and of other similar objects).
In other words, the cool phase may be regarded as
 a ``magnetosphere''  associated with the cD galaxy \citep[Paper I;][]{max97}.
The magnetic fields,
while working as a thermal insulator between the two phases,
 need not to be as strong as the equipartition value,
since the loop-interior plasma is confined 
primarily by external pressure from the hot phase.
This ``magnetosphere'' picture was already invoked 
successfully in  Paper I to explain the {\it ASCA} results.

Indeed, H$\alpha$-emitting filamentary features 
are optically observed in central regions of some clusters 
on a spatial scale of a few to few tens kpc \citep[e.g., ][]{perhalpha},
and the brightest of them can even be identified
in some {\it Chandra} X-ray images~\citep[e.g.][]{fabian03per}.
The best example is the recent {\it Chandra} result on M87 \citep{m87cxo},
which reveals rich filamentary structures
with a typical length of 10 to 50 kpc.
In the Centaurus cluster as well,
regions of strong magnetic fields are reported 
to be associated with filamentary structures
that emit H$\alpha$ photons and enhanced soft X-rays \citep{taylor07}.
While these filaments could be limb-brightened edges of bubbles
blown by the active galactic nucleus of M87 \citep{m87cxo},
they can alternatively  trace magnetic structures such as assumed there.

In addition to these  X-ray and radio results,
magnetohydrodynamic numerical simulations by \cite{asai07},
including both the radiative cooling and the field-alighned heat conduction,
show the emergence of low-temperature regions
along the loop-shaped magnetic field lines.
From these arguments,
we believe we have found a viable answer to
the issue (i) raised at the beginning of the present section.

\subsubsection{The Rosner-Tucker-Viana mechanism
\label{subsub:rtv}
}

We may assume that the cluster core region
is fed with a constant heating lumiosity $H$,
via some mechanism to be specified later in \S\ref{sub:heating}.
In such a case,
the volume heating rate will be generally
proportional to the local ICM density,
while the volume emissivity is obviously
proportional to the density squared.
Therefore, denser regions would preferentially cool and become even denser,
leading to a thermal instability: this is the issue (iii).
If, however, adopting the magnetosphere picture,
the cool phase can be thermally stabilized 
by a built-in  feedback mechanism,
hereafter called RTV mechanism,
originally developed  by \cite{rtv78} 
to explain quiet solar coronae.

The RTV mechanism holds for a thin magnetic flux tube
hydrostatically immersed in an external pressure,
with the two loop footpoints  anchored to a cool matter reservoir.
The loop-interior plasma is assumed to be heated with a constant luminosity $H$,
either uniformly along the loop or at the loop summit.
The deposited heat $H$ is assumed to flow toward the  footpoints 
due to loop-aligned conduction of eq.(\ref{eq:conduction}), 
and ultimately radiated away from various heights of the loop.
Suppose that $H$ was initially balanced by the radiative cooling,
but some perturbation caused $H$ to decrease.
Then, the loop loses its internal pressure,
and becomes thinner under the external compression,
thus reducing the conductive heat flow along it.
At the same time, 
some portion of the plasma flows into the matter reservoir, 
and the consequent decrease in the emission integral
reduces the radiative energy loss.
These concordant  effects bring the loop into 
a new steady state corresponding to a lower value of $H$.
Thus, the RTV mechanism thermally  stabilizes the cool phase,
and gives an answer to  the issue (iii).

Following the original work by \cite{rtv78},
\cite{rtv95,rtv96} used the {\it Yohkoh} spacecraft
to examin the RTV mechanism.
They confirmed
that it  is  likely to be actually working  in quiescent solar coronae,
which keep emitting X-rays in an apparently steady manner for a time much 
longer than the nominal cooling time ($\sim 30$ minutes).

\subsubsection{Cool-phase temperature 
\label{subsub:explainTc}
}

The RTV mechanism further allows us to answer (iv),
namely,  to  explain the absolute values  of $T_{\rm c}$.
Let us assume that the matter reservoir at the footpoints
is much cooler than the loop interior,
so that heat inflow to the reservoir can be neglected. 
Under such boundary conditions,
the temperature of the loop-confined plasma 
is shown to take the maximum value of
\begin{equation}
T_{\rm max} = (1.1-1.4) \times 10^3 (p_{\rm ext} l)^{1/3} ~~({\rm K})
\label{eq:rtv1}
\end{equation}
at the loop top \citep{rtv78,rtv95}.
Here, $l$ (cm) is the loop half length,
and $p_{\rm ext}$ is the loop-confining external pressure in units of dyn cm$^{-2}$.
This is so-called RTV temperature scaling,
which is actually confirmed in quiescent solar coronae \citep{rtv95,rtv96}.
The range of the coefficient, 1.1 to 1.4,
reflects differences in the assumed heat input distribution along the loop.
The index of scaling in eq.(\ref{eq:rtv1}) 
depends weakly on the form of the cooling function \citep{rtv96};
the particular value of 1/3 is valid over
a temperature range of 0.5--2 keV and near-solar abundances.

Most importantly,  
$T_{\rm max}$  in eq.(\ref{eq:rtv1})  is determined solely by the product $p_{\rm ext} l$,
without depending  on  $H$.
The loop-ineterior density $n_{\rm e}$ becomes also independent of $H$,
because  $p_{\rm ext} \propto n_{\rm e} T_{\rm max}$  is assumed to be constant.
Changes in $H$ affect only the loop cross section $S$, 
and hence the cool-phase X-ray luminosity
which is by definition equal to $H$.
More specifically,  eq.(\ref{eq:conduction}) and eq.(\ref{eq:rtv1}),
together with $ H \propto \kappa T_{\rm max} S/l$,
can be combined into another scaling relation as
\begin{equation}
S \propto H \: p_{\rm ext} ^{-7/6}\:  l^{-1/6}~~.
\label{eq:cross_section}
\end{equation}
Thus, the loop becomes thicker under  higher values of $H$.
The loop also gets thicker 
if we decrease  $p_{\rm ext}$  while keeping $H$ constant, 
but in this case, $T_{\rm max}$ automatically decreases via eq.(\ref{eq:rtv1}), 
so as to reduce the  conductive heat flux 
(per unit cross section) $\propto \kappa T_{\rm max}/l$
and hence  keep $H$ unchanged.

In the present case of Centaurus, 
we may identify $p_{\rm ext}$ with the hot-phase pressure in the core region,
namely $p_{\rm h} =2 n_{\rm h} k T_{\rm h} \sim 1 \times 10^{-10}$ dyn~cm$^{-2}$
with $n_{\rm h} \sim 1 \times 10^{-2}$ cm$^{-3}$ \citep{ikebe99},
and rewrite  eq.(\ref{eq:rtv1})  as
\begin{equation}
T_{\rm max} = (2.0-2.5) \left[
                    \left( \frac{p_{\rm h}}{1 \times 10^{-10}} \right)
                    \left( \frac{l}{\rm 30 kpc} \right)
               \right]^{1/3} ~~{\rm keV}~.
\label{eq:rtv2}
\end{equation}
Here, we assumed that loop semi-lenght to be
comparable to the narrower core radius, $\sim 30$ kpc,
of the double-$\beta$ modeling for the gravitational potential
in the Centaurus cluster \citep{ikebe99}.
This is because  each magnetic loop,
with its  interior having a higher density
than the surrounding hot phase,
may not become taller than the flat core radius
of the central gravitational potential.
The assumed length is comparable to those
of the soft X-ray filaments in M87 \citep{m87cxo}.

When  such a loop is observed as an integrated entity,
we expect to measure an X-ray temperature averaged over the loop length.
Assuming the footpoint temperature to be  0.7 keV (\S\ref{sub:0.7keV}),
a simple numerical solution to the RTV equation predicts 
this average to be  $\sim 0.7 \; T_{\rm max}$,
or $1.4-1.8$ keV from eq.(\ref{eq:rtv2}).
This is in an excellent agreement with $T_{\rm c}$
actually measured in the Centaurus cluster,
giving an answer to the issue (iv).

In reality, the loop length may well scatter,
with  longer loops tending to be  hotter according to eq.(\ref{eq:rtv2}).
Then, we expect the spherically averaged loop temperature 
to slightly increase outward, 
because  each loop is hottest at its top region,
and a larger radii would sample longer loops.
This agrees with what we actually observe in Figure~\ref{fig:2prad}.

\subsubsection{The $T_{\rm c}$ vs. $T_{\rm c}$ scaling
\label{subsub:Tc.vs.Th}}

Taking it for granted that $T_{\rm c}$ 
can be identified with $\sim 0.7 \; T_{\rm max}$,
let us consider the issue (iv), namely, 
to explain the scaling of eq.(\ref{eq:TcTh}).
The luminosity of the hot phase in a cluster is empirically 
known to scale with its temperature as
$L_{\rm h} \propto T_{\rm h}^{3.0}$ \citep[e.g.,][]{lt97}.
On the other hand, 
we generally have
 $L_{\rm h} \propto n_{\rm h}^2 R_{\rm h}^3T_{\rm h}^{1/2}$,
where $R_{\rm h}$ is a typical radius of the hot phase.
Elimination of $L_{\rm h}$ from these two equations gives
$n_{\rm h} \propto T_{\rm h}^{5/4} R_{\rm h}^{-3/2}$,
and hence  $p_{\rm h} \propto n_{\rm h} T_{\rm h} 
            \propto T_{\rm h}^{9/4} R_{\rm h}^{-3/2}$.
Substituting this into eq.(\ref{eq:rtv2}),
we obtain
\begin{equation}
 T_{\rm c} \propto T_{\rm h}^{3/4}  R_{\rm h}^{-1/2} l_{\rm h}^{1/3}~.
\label{eq:rtv3}
\end{equation}
Assuming that $R_{\rm h}^{-1/2} l_{\rm h}^{1/3}$ is relatively constant,
$T_{\rm c}$ indeed becomes roughly proportional to $T_{\rm h}$,
providing an explanation to eq.(\ref{eq:TcTh}).

As shown so far,
the RTV mechanism can account for the stable coexistence
of the cool and hot phases in the cluster core region.
The scenario is independent of the amount of available heat input,
because a lower value of $H$ simply reduces the cool-phase luminosity.
Furthermore, the RTV scaling can quantitatively
explain the absolute values of $T_{\rm c}$,
as well as its (near) proportionality to $T_{\rm h}$  
expressed by eq.(\ref{eq:TcTh}).
Thus, three of the four issues
raised at the beginning of \S \ref{sub:magnetosphere} have been answered.
We therefore propose the  magnetosphere model,
incorporating the RTV mechanism,
as a promising hypothesis to explain the plasma physics
in the central regions of cD clusters in general.

\subsubsection{Galaxy vs. ICM interaction
\label{subsub:interaction}}

To complete our new viewpoint, we may ask ourselves 
why  the bright cool emission is observed predominantly 
around cD  (or XD) galaxies
\citep[Figure 1 of Paper I]{jones_forman84},
including  NGC~4649.
Conversly, non-cD galaxies, like NGC~4472 in the Virgo cluster,
generally lack such bright cool X-ray emission,
even if they are optically as luminous as their cD counterparts.

An outstanding characteristic of a cD galaxy is
that it is nearly in a  standstill at the bottom of the gravitational potential.
Then, we may speculate 
that a substantial magnetosphere can develop 
around a galaxy in a cluster environment
only when it is at rest with respect to the ICM.
When a galaxy is moving through the ICM,
its magnetosphere will be disrupted by the ICM  ram pressure,
which  can be one to two orders of magnitude higher 
than the magnetic pressure of the assumed magnetosphere  (Paper I).
Such strong interactions are actually  revealed
by {\it Chandra} observations in nearby clusters \citep[e.g.,][]{n1404cxo,iizukaphd}.
Furthermore, optical observations of distant clusters
reveal numerous blue spiral galaxies 
with distorted morphology \citep[e.g.,][]{vandokkum01},
which we suggest to be interacting with the ICM.

\subsection{Possible Heating Mechanisms}
\label{sub:heating}

Having answered four out of the 5 questions
raised at the beginning of \S\ref{sub:magnetosphere},
the only remaining issue is (ii), i.e., to seek for possible heating 
mechanisms to sustain the cool-phase luminosity.
In the case of the Centaurus cluster,
the target value is $H=L_{\rm c}^{\rm bol}= 1.0 \times 10^{43}$ erg s$^{-1}$.
After Makishima (1997), Paper I, and the discussion conducted so far,
we retain our magnetosphere viewpoint.

The argument in \S\ref{subsub:interaction} indicates 
that  moving galaxies in a cluster interact with the ICM,
and  transfer their  kinetic energies to the ICM.
This process is estimated to proceed at a rate of  
\begin{equation}
  L_{\rm int} \sim N  n m_{\rm p} v^3 \pi R_{\rm int}^2
\label{eq:interaction}
\end{equation}
\citep{sarazin},
where $N$ is the number of moving galaxies,
$n$ is the average ICM number density, 
$v$ is the average galaxy velocity,
and $R_{\rm int}$ is the interaction radius for each galaxy.
We may use $v=1,010$ km s$^{-1}$ 
which is $\sqrt{3}$ times the velocity dispersion of 
the Centaurus cluster (586~km s$^{-1}$) after \cite{cenopt86},
$n \sim 0.8 \times 10^{-3}$ cm$^{-3}$
which is an average over the central 300 kpc \citep{ikebe99},
and $N \sim 50$.
Furthermore,  a relatively large value of $R_{\rm int}$,
say, $\sim 5$ kpc, is suggested by
the argument  in \S\ref{subsub:interaction},
and by some X-ray observations \citep{iizukaphd}.
Through eq.(\ref{eq:interaction}), these numbers yield
$L_{\rm int} \sim 4.5 \times 10^{43}$ erg s$^{-1}$,
which  well exceeds $L_{\rm c}^{\rm bol} \sim 1\times 10^{43}$ erg s$^{-1}$.
Therefore, the galaxy-ICM interaction becomes a
promising candidate for the ICM heating,
and for the suppression of CFs.
These effects have been reproduced
successfully by numerical simulations \citep{asai07}.

How large is the kinetic energy  $E_{\rm kin}$ 
stored in the moving galaxies?
The stellar mass of the Centaurus cluster within the central 300 kpc,
but excluding the cD galaxy,
is estimated as $1.5 \times 10^{12}~M_\odot$ \citep{ikebe99}
after correcting for the different values of the Hubble constant.
We  have then $E_{\rm kin}\sim 1.5  \times 10^{61}$ ergs,
which is likely to be a lower limit
because the employed galaxy mass assumes 
a conservative mass-to-light ratio of 8.
As a result, the time scale of the energy tansfer
from the moving galaxies to the ICM is estimated as
$E_{\rm kin}/L_{\rm int} \sim 1.1 \times 10^{10}$ yr.
Therefore, the kinetic energy in the moving galaxies is large 
enough to sustain $L_{\rm int}$ over the Hubble time.

The galaxy-to-ICM energy transfer  is expected
to cause galaxies to gradually fall  to the potential center.
There are several pieces of observational evidence 
suggesting that this effect is actually taking place.
In fact, the stellar-mass distributions in many clusters 
are known to be much more centrally peaked than that of the ICM.
Furthermore,  clear central decreases in ``iron-mass to light ratio'',
observed from a fair number of clusters and groups
\citep[Paper~I;][]{kawahard, n1550},
suggest that galaxies used to be distributed  more widely than  today,
and polluted the ICM with metals out to the cluster edges.

The energy released  by the moving galaxies may be 
stored  once in the ICM in the form of turbulence and bulk flows.
However, according to the  {\it Suzaku} observation  \citep{ota07},
these kinetic energies in the  ICM  of the Centaurus cluster,
as estimated from the width and position-dependent centroid shifts of the Fe-K line,
cannot largely exceed its thermal energy.
Therefore, the turbulent and bulk-flow energies 
must be dissipated and thermalized efficiently  in the ICM,
with a rate  comparable to $L_{\rm int}$.
If  a fair fraction of
the dissipated energy  is deposited onto the magnetosphere,
the available heating luminosity of the cool phase becomes
of the order of $H \sim 1 \times 10^{43}$ erg s$^{-1}$
which  can just  sustain  the observed $L_{\rm c}^{\rm bol}$.
We would rather say that $L_{\rm c}^{\rm bol}$ 
is self-adjusted, via the RTV mechanism,
to match the predetermined energy dissipation rate.
The remaining portion of $L_{\rm int}$ will  heat the hot phase,
to make it spatially more extended than the total gravitating mass.

The actual dissipation of turbulence and bulk-flow energies
may take place through magnetohydrodynamic processes.
For example,  moving galaxies will pick up 
magnetic field lines in the ICM and stretch them,
exciting Alfv\'en waves and causing field lines to reconnect.
When the magnetic reconnection occurs between two closed magnetic loops,
the released energy will be spent in heating the cool phase.
That between closed and open field lines will expel a small 
potion of the metal-rich cool-phase plasma into the hot phase.
The ejected plasmoid will become quickly isothermal with the  hot phase,
in terms of the electron temperature via eq.(\ref{eq:conduction}),
as well as  in terms of the ion ionization temperature
 \citep[in a few times $10^7$ yr; ][]{masai84}.
The resulting highly ionized heavy ions will remain there 
due to their slow diffusion \citep{ezawa97},
and increase the metallicity of the hot phase.
This can explain why the central metallicity increase is observed
in both phases.

In addition to the above mechanism,
there can be another ICM heating mechanism 
localized around each cD galaxy;
namely, a shrink of its self-gravitating core.
Since this has already been discussed in Paper I,
we simply mention
that this mechanism can also account for a
heating luminosity of $\sim 1 \times 10^{43}$ erg s$^{-1}$.

\section{CONCLUSION}

Analyzing the {\it XMM-Newton} EPIC and RGS data  
of the Centaurus cluster,
we have obtained  the following results.
\begin{enumerate}
\vspace*{-2mm}
\setlength{\itemsep}{0mm}
\item 
The ICM in the central $\sim 75$ kpc 
can be described  better by the two-phase view
(plus the 0.7 keV component at $<12$ kpc) 
than by the  single-phase picture.
\item
There is no evidence of cooling flows.
\item
The iron and silicon abundances increases significantly 
in the central region, while that of oxygen is radially constant.
\item 
The previously reported excess X-ray absorption disappears
when the central oxygen deficit is  considered.
\item
The overall results agree with those  by \cite{ikebe99}
who  used  {\it ASCA} and {\it ROSAT}.
\end{enumerate}

To explain these observations,
we have developed a working hypothesis (after Paper~I),
in which the cool phase is thought to constitute 
a magnetosphere associated with the cD galaxy.
Incorporating the Rosner-Tucker-Vaiana mechanism,
it can account for the following essential features of the Centaurus cluster,
as well as of  similar cD clusters.
\begin{enumerate}
\setlength{\itemsep}{-1mm}
\vspace*{-2mm}
\item The hot and cool ICM phases co-exsit in the cluster core region.
\item The ICM is provided with a  heating luminosity, 
        which is high enough to sustain
         the X-ray emission against the radiative cooling.
\item  The cool phase is kept thermally stable.
\item The cool phase has a typical temperature of $\sim 2$ keV,
        with a mild outward increase.
\item A good proportionality holds between the cool and hot temperatures.
\item The ICM metallicity becomes enhanced in both phases toward the center.
\end{enumerate}
The present results thus provide a new insight into the physics of clusters of galaxies.

\bigskip\noindent
Acknowledgements:
This work was supported in part by  Grant-in-Aid for Scientific research (S)
(Japanese Ministry of Education,  Culture,  Sports, Scince \& Technology)
 \#18104004  on``Study of Interactions between Galaxies and  Intra-Cluster Plasmas",
as well as by Special Research Project for Basic Science
(Institute of Physical and Chemical Research) on 
``Investigation of Spontaneously Evolving Systems".

\clearpage

\section*{Apendix A: The Deprojection Matirx}
This Appendix is meant to provide numerical  details
of the deprojection process described in \S\ref{subsub:depro}.
Let us consider 11 annular regions, of which the outer radii are
at $0'.5$, $1'.0$, $1'.5$, $2'$, $3'$, $4'$, $5'$, $6'$, $8'$, $10'$, and $12'$,
and the corresponding 11 shell regions.
These divisions are the same as used in
Figures~\ref{fig:rawpspec}, \ref{fig:1prad}, and \ref{fig:2prad},
except that the innermost 4 regions therein are here re-arranged into 
two regions just for  clearer presentation.
The reconstructed  ``Shell C'' spectrum,  
namely $C \equiv \Sigma_{j=3}^7 S_j$,
is not affected by this simplified treatment of inner annuli/shells.
Let $(A_1, A_2, ..., A_{11})$ be the projected spectra from the 11 annular regions
at a given energy,
while $(S_1, S_2, ..., S_{11})$ those from the 11 shells.
Note $A_{11}$ has the outer boundary at $12'$,
while $S_{11}$ includes emission outside that.
Then, let ${\cal M}_{i,j}$ denote the projection matrix 
which converts $\{ S_i \}$ into $\{ A_j \}$.
In the present geometrical setting, it is numerically given as
\begin{eqnarray}
\begin{footnotesize}
\left(
     \begin{array}{ccccccccccc}
        A_1\\
        A_2\\
        A_3\\
        A_4\\
        A_5\\
        A_6\\
        A_7\\
        A_8\\
        A_9\\
        A_{10}\\
        A_{11}
     \end{array}
\right) = 
\left(
    \begin{array}{ccccccccccc}
1.000 &0.258 &0.083 &0.041 &0.020 &0.010 &0.006 &0.004 &0.003 &0.002 & 0.004\\
0     &0.742 &0.329 &0.137 &0.063 &0.031 &0.019 &0.013 &0.008 &0.005 & 0.011\\
0     &0        &0.588 &0.321 &0.116 &0.055 &0.032 &0.021 &0.013 & 0.008 &0.018\\
0     &0        &0        &0.501 &0.213 &0.083 &0.047 &0.031 &0.018 & 0.011 &0.026\\
0     &0        &0        &0        &0.588 &0.321 &0.151 &0.093 &0.055 & 0.032 &0.074\\
0     &0        &0        &0        &0        &0.501 &0.303 &0.152 &0.083 & 0.047 &0.105\\
0     &0        &0        &0        &0        &0        &0.443 &0.285 &0.122 & 0.064 &0.137\\
0     &0        &0        &0        &0        &0        &0        &0.401 &0.199 & 0.086 &0.174\\
0     &0        &0        &0        &0        &0        &0        &0        &0.501& 0.303 &0.477\\
0     &0        &0        &0        &0        &0        &0        &0        &0        &0.443  &0.724\\
0     &0        &0        &0        &0        &0        &0        &0        &0        &0         &1.113\\
 \end{array}
\right)
\left(
     \begin{array}{ccccccccccc}
        S_1\\
        S_2\\
        S_3\\
        S_4\\
        S_5\\
        S_6\\
        S_7\\
        S_8\\
        S_9\\
        S_{10}\\
        S_{11}
     \end{array}
\right)~.
\label{eq:projnum}
\end{footnotesize}
\end{eqnarray}
On the assumption that the emission is uniform within each individual shell,
these matrix elements are uniquely determined by the geometry alone,
without any dependence on the emission model or instrumental effects.
The only exception is the last column, namely ${\cal M}_{i,11}$ ($i=1,2,..,11$),
which assumes the emission outside $12'$ as explained in \S \ref{subsub:depro}.
Although the matrix elements are shown with only 3 digits below decimal points,
this is again for simplicity;
the calculation is preformed with double precision. 

By inverting this matrix, we can calculate the deprojection matrix $\{ {\cal D}_{i,j} \}$
defined in eq.(\ref{eq:deprojection}).
Numerically, it is calculated as
\begin{scriptsize}
\begin{eqnarray}
\left(
     \begin{array}{ccccccccccc}
        S_1\\
        S_2\\
        S_3\\
        S_4\\
        S_5\\
        S_6\\
        S_7\\
        S_8\\
        S_9\\
        S_{10}\\
        S_{11}
     \end{array}
\right) = 
\left(
    \begin{array}{ccccccccccc}
1.00 &-0.347 &0.054 &-0.022 & 0.000 &-0.001 &-0.000 &-0.000 &-0.000 &-0.000 & 0.000\\
0     &1.347 &-0.753 & 0.113  &-0.035 & 0.002 &-0.004 &-0.001 &-0.001 &-0.000  & -0.000\\
0     &0        &1.700  &-1.089  &0.058  &-0.042 & 0.001 &-0.005 &-0.002 &-0.001 &-0.001\\
0     &0        &0        &1.998    &-0.722 & 0.132 &-0.057 &0.006  &-0.005 &-0.001 &-0.002\\
0     &0        &0        &0          &1.700   &-1.089 &0.167  &-0.101  &-0.006 &-0.009 &-0.007\\
0     &0        &0        &0          &0          & 1.998 &-1.367 &0.215  &-0.083 & 0.002 &-0.019\\
0     &0        &0        &0          &0          &0         & 2.259 &-1.608 &0.090 & -0.076 &-0.017\\
0     &0        &0        &0          &0          &0         &0        & 2.494 &-0.992 & 0.192 &-0.090\\
0     &0        &0        &0          &0          &0         &0        &0        &1.998& -1.367 &0.034\\
0     &0        &0        &0          &0          &0         &0        &0        &0        &2.259  &-1.471\\
0     &0        &0        &0          &0          &0         &0        &0        &0        &0         &0.899\\
 \end{array}
\right)
\left(
     \begin{array}{ccccccccccc}
        A_1\\
        A_2\\
        A_3\\
        A_4\\
        A_5\\
        A_6\\
        A_7\\
        A_8\\
        A_9\\
        A_{10}\\
        A_{11}
     \end{array}
\right)~.
\label{eq:depronum}
\end{eqnarray}
\end{scriptsize}

From eq.(\ref{eq:depronum}), 
the Shell C spectrum $C$ at each energy can be expressed as
\begin{equation}
C = (1.700 A_3 + 0.909 A_4 + 1.036 A_5 + 0.999 A_6 + 1.003 A_7)\\
 - (1.493 A_8 + 0.005 A_9 + 0.085A_{10}+  0.045 A_{11}) .
 \label{eq:ShellCnum}
\end{equation}
Thus, the outermost annulus $A_{11}$,
which is most subject to the background uncertainty
as well as to that of the emission outside $R=12'$,
contributes {\it geometrically} only less than 5\% to the Shell C spectrum,
compared to the individual shells $A_3$ through $A_7$.
In addition, the flux in $A_{11}$ is  $\sim 3$ to $\sim 30$ times lower
than those in the 3rd to 7th annuli (Figure~\ref{fig:rawpspec}).
Considering these, 
we estimate that  the emission from the $R>12'$ region
contributes {\it physically} by no more than $\sim 1\%$ to the Shell C spectrum.
Therefore, any systematic error associated with $A_{11}$
is concluded to be negligible (\S \ref{subsub:artifacts}).

Another exercise to be conducted using eq.(\ref{eq:depronum}) is 
the issue raised in \S~\ref{subsub:thick_shell_spec},
i.e., how the assumption of emission constancy within each thin shell
affects our Shell C spectrum.
As mentionned there, 
let us examine what happens if we purposely neglected
spectral differences between $A_5$ and $A_6$
(the 3rd and 4th thin shell constructing Shell C, respectively),
where the 1P properties change rather steeply with $R$.
Taking the Fe-K line equivalent width as an example,
we may write the fluxes  in the Fe-K line energy region as 
$A_5=B_5 (1+W_5)$ and $A_6=B_6 (1+W_6)$,
where $B$ represents continuum and $W$ the line equivalent width.
Using the difference $\delta W \equiv W_5 - W_6$
and the emission-measure-weighted mean 
$\bar{W} \equiv (B_5 W_5 + B_6 W_6)/(B_5+W_6)$,
we may rewrite as 
$A_5=B_5 \left[ 1+\bar{W}+B_6 \delta W/(B_5+B_6) \right]$ 
and 
$A_6=B_6 \left[1+\bar{W}-B_5 \delta W/(B_5+B_6)\right]$.
Substitution of this into eq.(\ref{eq:ShellCnum}) readily yields
\begin{equation}
C = C'  + (1.036 B_5 + 0.999 B_6)(1+\bar{W})
 + 0.037\alpha \;(B_5+B_6)  \delta W
 \label{eq:ShellCdiff}
\end{equation}
where $C'$ represents terms unrelated to $A_5$ or  $A_6$, 
while $\alpha \equiv B_5B_6/(B_5+B_6)^2 \leq 1/4$ is a numerical factor.
Thus, the negligence of  $\delta W$ between the two annular spectra
has only an effect of at most $\sim 0.01 \delta W$ in the Shell C spectrum.
The essence is that the deprojection procedure has mathematically
 ``differential" nature exaggerating small-scale features,
while this problem is largely removed by  constructing 
a thick shell wherein a process of integration is involved.
This is evidenced by eq.(\ref{eq:ShellCnum}),
where $C$ is close to a simple summation of $A_3$ through $A_7$,
except that $A_3$ must be given a 1.7 times larger weight
and $A_8$ (the shell just outside Shell C) must be subtracted
with  a relatively large weight of 1.5.

\section*{Apendix B: A Numerical Simulation of an 1P Cluster}

\begin{figure}\begin{center}
\epsscale{0.6}
\plotone{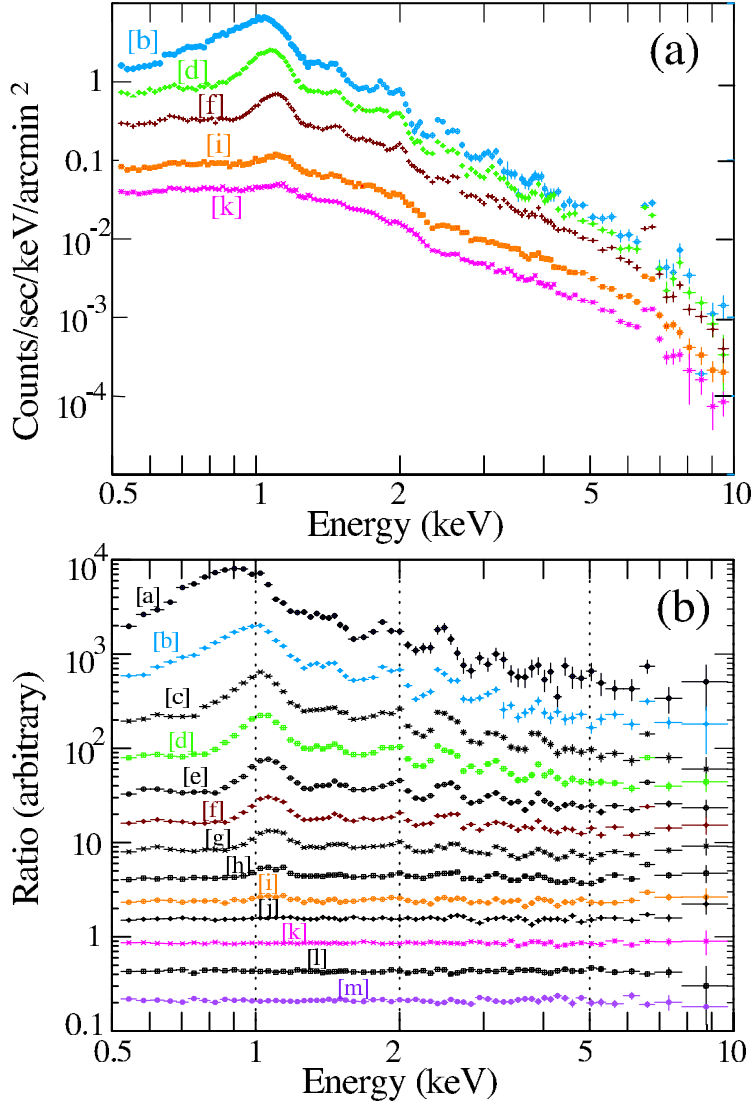}
\caption{
Background-subtracted EPIC-PN spectra of the Centaurus cluster
extracted from annular regions (panel a),
and their ratios to that averaged over the region of $5' < r < 12'$ (panel b).
Outer radii used to extract the spectra are
[a]$10''$, [b]$25''$, [c]$40''$, [d]$1'$, [e]$1'. 5$, [f]$2'$,
[g]$3'$, [h]$4'$, [i]$5'$, [j]$6'$, [k]$8'$, [l]$10'$, and [m]$12'$.
The spectra are not corrected for the vignetting.
In panel (a), 
only 5 out of the 13 spectra are shown for clarity.
The ratios in panel (b) are vertically offset
to avoid overlap.
}
\label{fig:rawpspec}
\end{center}
\end{figure}

\begin{figure}
\begin{center}
\epsscale{0.6}
\plotone{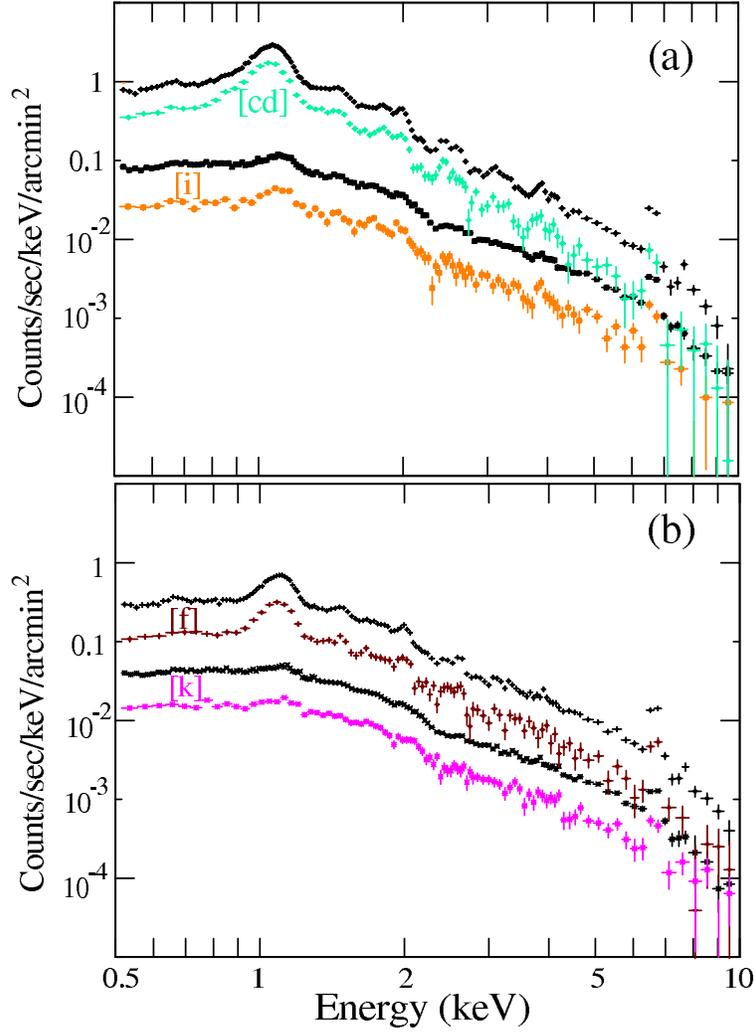}
\caption{
Deprojected PN spectra~(colors) from concentric thin shells,
compared with their non-deprojected counterparts (black).
The regions used are; 
[cd]$0'.5-1'$, [f]$1'.5-2'$, [i]$4'-5'$, and [k]$6'-8'$.
To avoid heavy overlaps,
a series of spectral pairs are shown on every other panel.
}
\label{fig:rawdspec}
\end{center}
\end{figure}

\begin{figure}\begin{center}
\epsscale{1.0}
\plotone{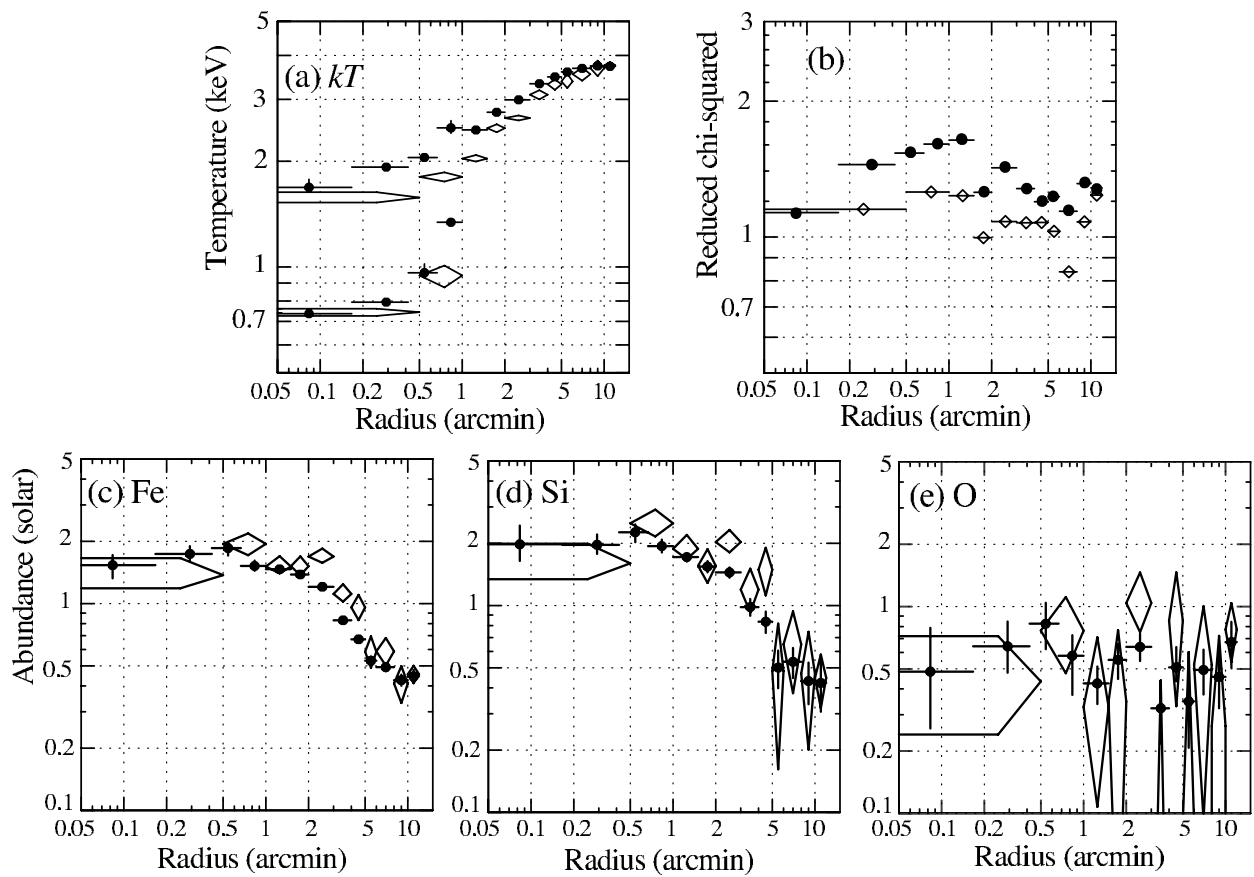}
\caption{
Radial profiles of 
(a) the temperature, (b) reduced chi-squared,
and the abundances of (c) Fe, (d) Si, and (e) O of the Centaurus cluster,
obtained with the 1P modeling.
A $\sim 0.7$~keV component is added within $r < 1'$.
Filled circles and open diamonds represent results from
the non-deprojected and deprojected spectra, respectively.
}
\label{fig:1prad}
\end{center}
\end{figure}

\begin{figure}\begin{center}
\epsscale{1.0}
\plotone{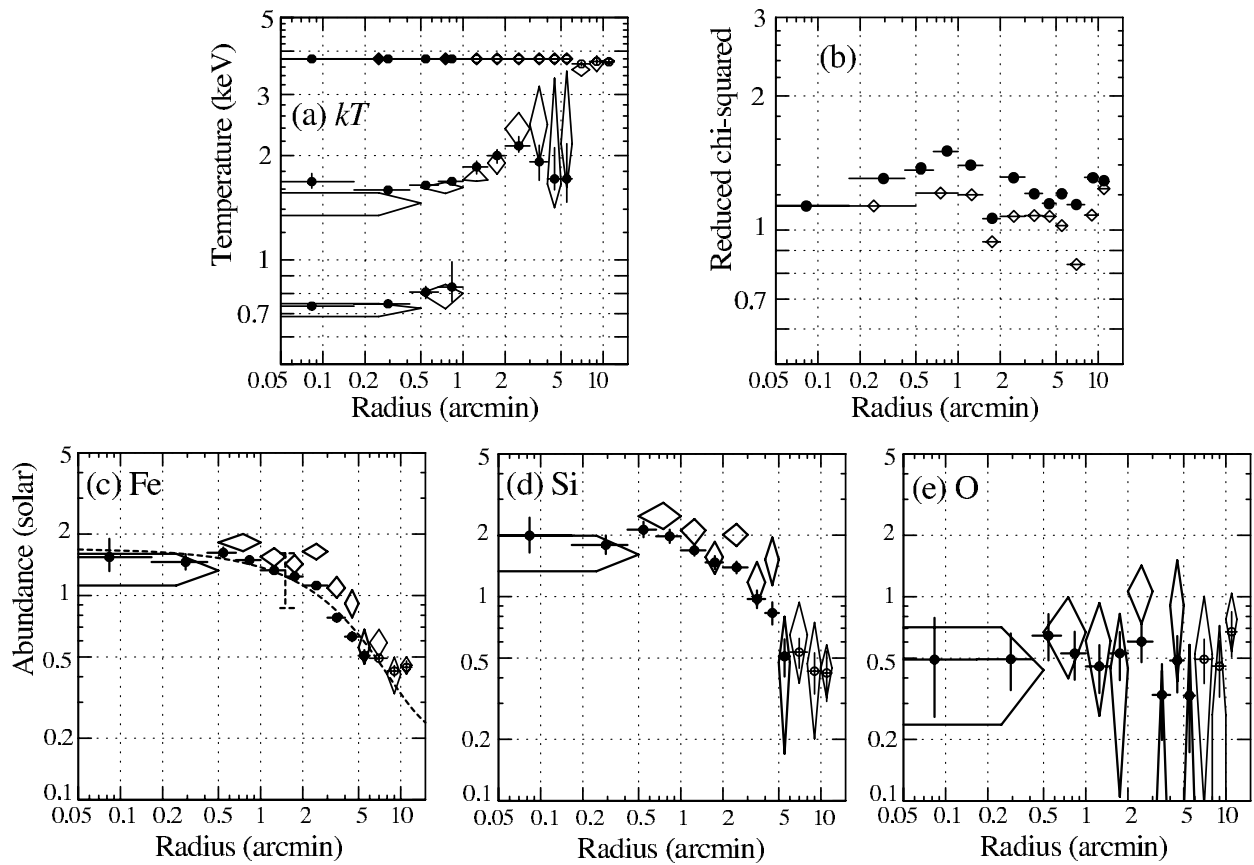}
\caption{
The same as Figure~\ref{fig:1prad}, but using the 2P model.
The results outside $6'$,
where the cool component is not required, 
are the same as those of Figure~\ref{fig:1prad}.
In panel (c),
the Fe abundance profile obtained with {\it ASCA}~\citep{ikebe99}
is shown in the dashed line,
together with its typical error size~(30~\%).
}
\label{fig:2prad}
\end{center}
\end{figure}

\begin{figure}\begin{center}
\epsscale{1.0}
\plotone{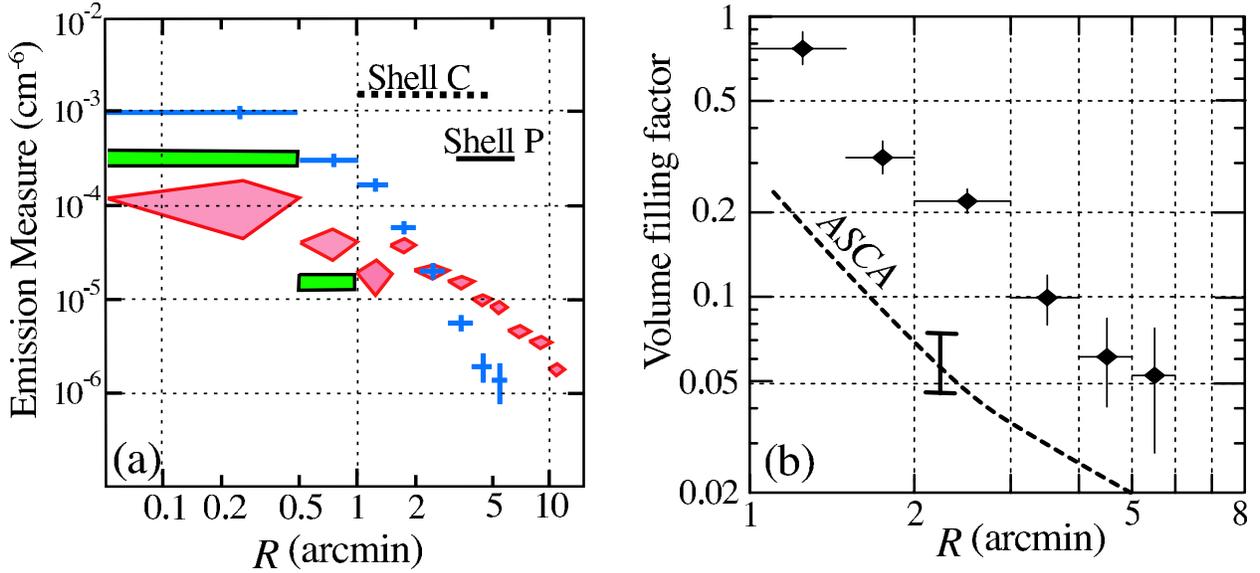}
\caption{
(a) Radial profiles of the emission measure per unit volume 
of the hot (red diamonds), cool (blue crosses),
and the 0.7 keV (green squares) components,
derived through the 2T fits to the deprojected EPIC spectra.
Dashed and dotted horizontal lines indicate 
the radial ranges covered by Shell~C and Shell~P, respectively.
(b) Radial profile of the volume filling factor of the cool component,
calculated via eq.(\ref{eq:filling_factor}).
That obtained with {\it ASCA}~\citep{ikebe99}
is also shown with a dashed line,
together with its typical error (25\%).}
\label{fig:vfill}
\end{center}
\end{figure}

\begin{figure}\begin{center}
\epsscale{1.0}
\plotone{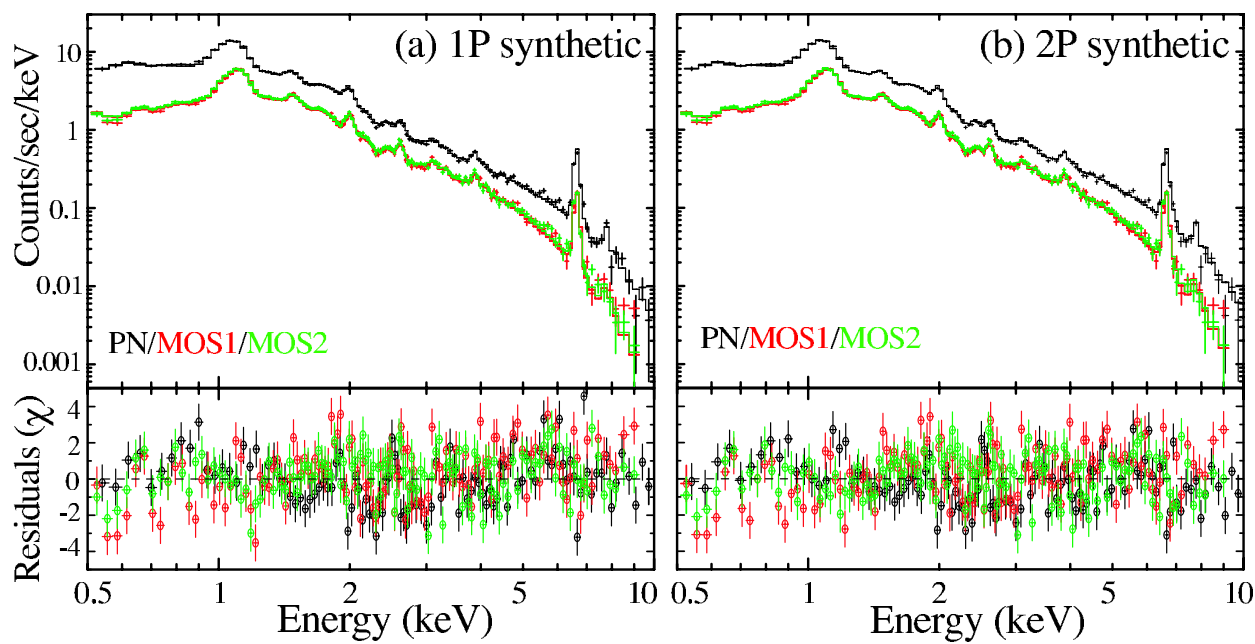}
\caption
{
(a) Actual spectra from the thick Shell~C,
compared with the synthetic 1P model spectra (solid histograms)
constructed from the best-fit 1P models 
to the constituent 5 thin-shell spectra (see text \S\ref{subsub:thick_shell_1P}).
The PN, MOS1, and MOS2 spectra are presented in black, red, 
and green, respectively.
(b) The same Shell~C spectra,
fitted jointly with a simple 2P model 
of free temperatures and free abundances (\S\ref{subsub:thick_shell_2P}).
The fit goodness is summarized in Table 1.
}
\label{fig:regCcmpr}
\end{center}
\end{figure}

\begin{figure}
\begin{center}
\epsscale{0.55}
\plotone{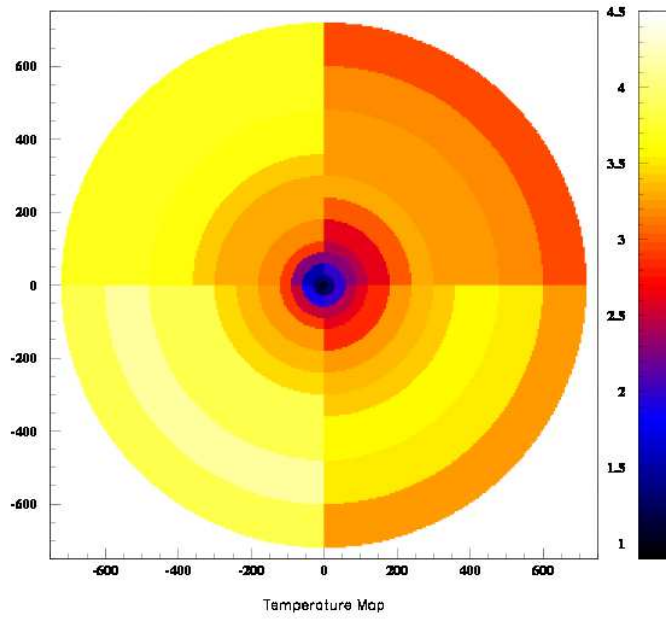}
\caption
{
A temperature map of the Centaurus cluster,
derived by analyzing the projected EPIC spectra
in four azimuthal sectors and the 12 radial intervales.
The MOS and PN spectra are fitted simultaneously 
by a 1P model (\S~\ref{subsub:ana1P}).
The map scale is in units of arcsec,
with north up and east to the left.
The temperature color scale is in the unit of keV.
}
\label{fig:Tmap}
\end{center}
\end{figure}

\begin{figure}
\begin{center}
\epsscale{0.5}
\plotone{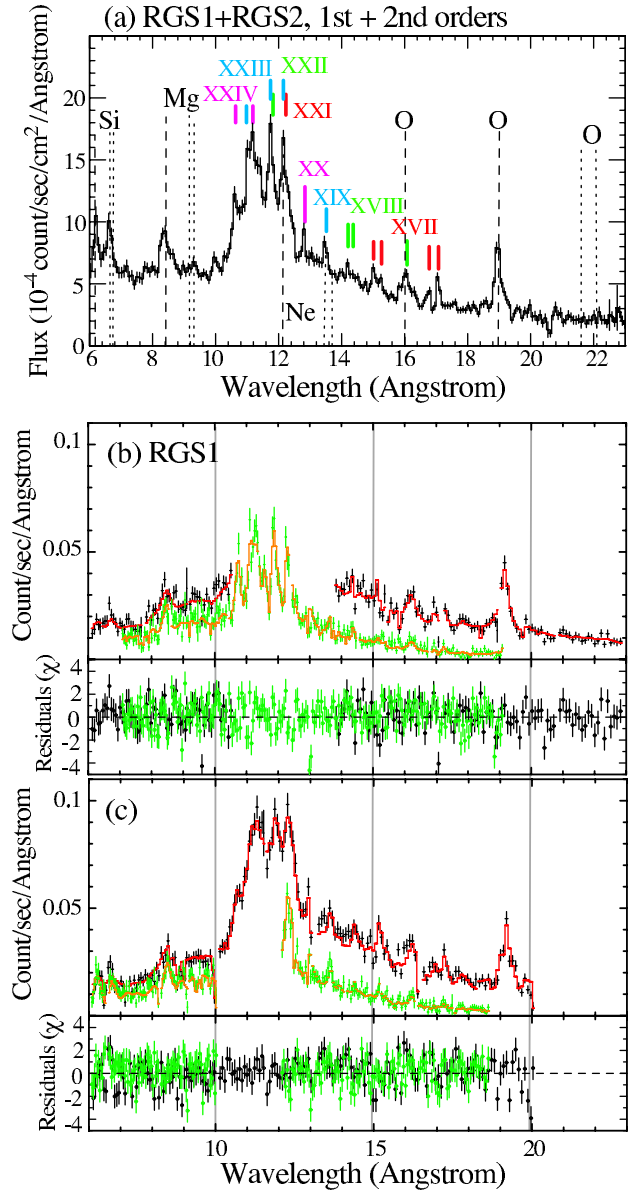}
\caption
{
(a) Background-subtracted RGS flux spectrum made by combining four spectra,
namely, the 1st and 2nd order spectra from the two RGS units.
The positions of the H-like and He-like K-lines of several major elements
are indicated with dashed and dotted lines, respectively.
Color tick marks indicate Fe-L lines of various ionization states.
(b) RGS1 spectra, fitted jointly with a 2P model.
The first order spectrum and its model prediction are
shown in black and red respectively,
whereas the 2nd order data and its model predictions are
in green and orange, respectively.
(c) The same as panel (b), but for RGS2.
}
\label{fig:rgs}
\end{center}
\end{figure}

\begin{figure}
\begin{center}
\epsscale{0.6}
\plotone{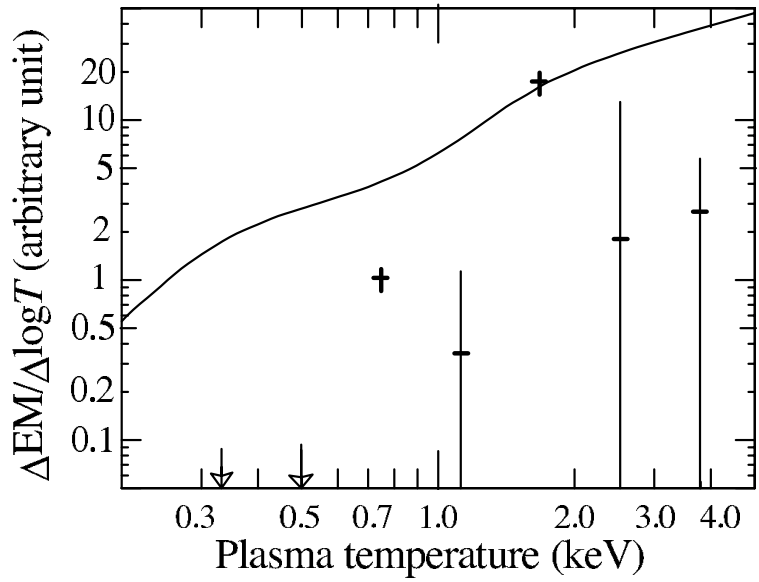}
\caption
{
Distribution of the normalizations of individual temperature components, 
obtained with the multi temperature fit to the RGS data.
The curve indicates the prediction by the isobaric cooling flow model,
for plasmas of 1 solar abundances \citep{isobaric}.
The vertical scale is arbitrary.
}
\label{fig:rgsdem}
\end{center}
\end{figure}

In order to assess the validity of our ``thick shell" method,
the following numerical simulation was performed.
We started from the best-fit 1P numerical models  (Figure~\ref{fig:1prad})
obtained for the 13 deprojected thin shell regions
which are defined in Figure~\ref{fig:rawpspec}.
Then, to represent the radial spectral changes more smoothly,
each of the 5 shells constituting Shell C (covering $r=1'-5'$) 
was subdivided into inner and outer halves,
and appropriate temperature and emission measure were assigned 
to the inner/outer pair.
These  model spectra, now comprising 18 shells,
were projected numerically onto the corresponding 18 annuli,
to form numerical spectral models for the 18 annular regions.
At this stage, the inner/outer pair were numerically recombined together.
The derived 13 model spectra were convolved with the EPIC energy responses,
and were given Poisson errors simulating the actual observation.
Then, in the same manner as the actual data,
we analyzed these 13 annular spectra  via deprojection.

\begin{figure}
\begin{center}
\epsscale{0.6}
\plotone{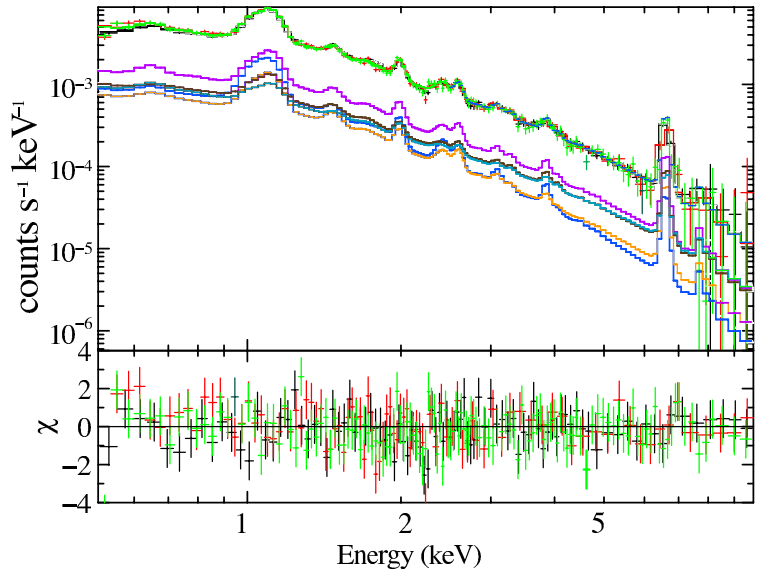}
\caption
{Spectra from Shell C of a simulated 1P cluster,
compared with the convolved numerical models
that are used to define the 5 constituent annular spectra
(indicated only for MOS1 with different colors).
The PN, MOS1, and MOS2 spectra are presented in black, red, 
and green, respectively.
Their appearance differs from that of Fig.~\ref{fig:regCcmpr},
simply because the energy-dependent effective areas,
which are once removed in deprojection,
are restored in Fig.~\ref{fig:regCcmpr} 
while not here for simplicity.
}
\label{fig:fake1p}
\end{center}
\end{figure}

Figure~\ref{fig:fake1p} shows the simulated Shell C spectra,
compared with the ``sysnthetic 1P spectra" 
introduced in \S \ref{subsub:thick_shell_1P}.
The comparison is fully satisfactory with $\chi^2/\nu=324.96/322=1.009$.
For reference, a 2P fit with two APEC components,
in the same manner as explained in  \S \ref{subsub:thick_shell_2P},
gave  $\chi^2/\nu=312.9/309=1.013$,
which  is similarly acceptable.
In other words, the 1P and 2P interpretations degenerate
unlike the case of the actual data.
Furthermore, unlike the 2P fit to the actual Shell C spectra
which revealed the $T_{\rm h}\sim 4$ keV component,
the 2P fit to the simulated data yielded
 $T_{\rm c}=2.00^{+0.14}_{-0.35}$ keV and $T_{\rm h}=3.11 ^{+0.09}_{-0.16}$ keV,
which are both consistent with the assumed 1P temperature range (2.03 to 3.30 keV).



\clearpage
\begin{table}[hbt]
\caption{Systematic uncertainties (1 $\sigma$) assigned to the EPIC background data.}
\bigskip
\begin{tabular}{cccc}
\hline \hline 
 & band (keV)  & errors & note  \\
\hline
PN  &$ < 1.35 $  &   8\%   & diffuse X-rays  \\
  &1.35--1.6    &   8\%  & Al line                \\
  &1.6--7.3     &  3\%    &\\
  & 7.3--9.2    & 8\%     & Ni, Cu, and Zn lines \\
  & $> 9.2$    &  3\% \\
\hline 
MOS  &  $< 1.35$   & 8\% &  diffuse X-rays\\
 &  1.38--1.85 & 8\% & Al and Si lines\\
 &  $> 1.85$   & 3\% \\
\hline 
\end{tabular}
\label{tbl:syserror}
\end{table}

\begin{table}[hbt]
\caption{Summary of the fit goodness to the Shell~C spectra.}
\bigskip
\begin{tabular}{clccccc}
\hline\hline
Model & Condition    &$\chi^2$ & $\nu$ & $\chi^2/\nu$ & \S\S \\
 \hline
1P & synthetic, no adjust. & 739 & 378 & 1.96 & 4.2.2\\
   & norm. adjustted       & 724 & 373 & 1.94 & 4.2.2\\
   & revised sys. error    & 418 & 378 & 1.11 & 4.2.4\\
\hline 
2P & simple 2P fit         & 688 & 363 & 1.90 & 4.2.3\\
   & synthetic, no adjust. & 678 & 378 & 1.79 & 4.2.3\\
   & revised sys. error    & 381 & 378 & 1.01 & 4.2.4\\
\hline 
\end{tabular}
\label{tbl:chisquares}
\end{table}

\begin{table}
\caption{Results of the model fit to the RGS spectra.}
\label{tab:rgsfit}
\bigskip
\begin{tabular}{lcc}
\hline 
\hline 
  Parameter & Comp. 1   & Comp. 2           \\
\hline 
Temperature   & $1.70^{+0.05}_{-0.03}$ &  $0.77^{+0.02}_{-0.04}$ \\
Gaussian $\sigma$ 
              & $65''.8^{+5''.5}_{-5''.1}$  & $19''.6^{+7''.9}_{-6''.8}$\\
\hline 
\multicolumn{3}{l}{Abundances (solar)} \\
~~~~O  & \multicolumn{2}{c}{$0.39 \pm 0.04$       } \\
~~~~Ne & \multicolumn{2}{c}{$0.60 \pm 0.12$       } \\
~~~~Mg & \multicolumn{2}{c}{$0.75^{+0.12}_{-0.13}$} \\
~~~~Si & \multicolumn{2}{c}{$1.09^{+0.13}_{-0.12}$} \\ 
~~~~Fe & \multicolumn{2}{c}{$0.75^{+0.04}_{-0.05}$} \\
~~~~Ni & \multicolumn{2}{c}{$1.01^{+0.26}_{-0.25}$} \\
\hline 
$\chi^2$/d.o.f. & \multicolumn{2}{c}{985/727} \\
\hline 
\end{tabular}
\end{table}

\end{document}